\documentclass[
aps,
pra,
,amsmath
,amssymb
,amsfonts
,twocolumn
,tightenlines
,superscriptaddress
]{revtex4-2}
\usepackage[dvipsnames]{xcolor}

\usepackage{graphicx}
\usepackage{dcolumn}
\usepackage{bm}
\usepackage{hyperref}
\hypersetup{colorlinks = true, linkcolor = [rgb]{0.19411,0.51882,0.667058}, urlcolor = [rgb]{0.125490,0.29542,0.1647058}, citecolor = [rgb]{0.75882,0.37411,0.14117}}

\usepackage{setspace}

\usepackage{amssymb}

\usepackage{braket}

\usepackage[normalem]{ulem}

\graphicspath{{../images/}}
\usepackage[toc,page]{appendix}

\newcommand{\nn}{\nonumber \\}

\usepackage{braket}

\def\>{\rangle}
\def\<{\langle}

\usepackage{color}

\usepackage{subfigure}

\usepackage{graphicx}
\usepackage{bm}
\usepackage{amsmath}
\usepackage{amssymb}

\begin{document}

\title{Error filtration for quantum sensing via interferometry}

\author{Zixin Huang}
\email{zixin.huang@mq.edu.au}
\affiliation{School of Mathematical and Physical Sciences, Macquarie University, NSW 2109, Australia}
\affiliation{Centre for Quantum Software and Information, Faculty of Engineering and IT, University of Technology, Sydney, Australia}

\author{Cosmo Lupo}
\email{cosmo.lupo@poliba.it}
\affiliation{Dipartimento  Interateneo di Fisica, Politecnico \& Universit\`a di Bari, 70126, Bari, Italy}
\affiliation{INFN, Sezione di Bari, 70126 Bari, Italy}

\date{\today}

\begin{abstract}
Dephasing is a main noise mechanism that afflicts quantum information, it reduces visibility, and destroys coherence and entanglement. 
Therefore, it must be reduced, mitigated, and if possible corrected, to allow for demonstration of quantum advantage in any application of quantum technology, from computing to sensing and communications.
Here we discuss a hardware scheme of error filtration to mitigate the effects of dephasing in optical quantum metrology.
The scheme uses only passive linear optics and ancillary vacuum modes, without need of single-photon sources or entanglement. 
It exploits constructive and destructive interference to partially cancel the detrimental effects of statistically independent sources of dephasing. 
We apply this scheme to preserve coherent states and to phase-stabilize stellar interferometry, showing that a significant improvement can be obtained by using only a few ancillary modes.
\end{abstract}

\maketitle

\section{Introduction}

When information is encoded into quantum states, there are tasks in computing~\cite{Shora,Grovera,QChem,gauge,QHEP} sensing~\cite{Dowling,JPD,PhysRevLett.109.070503,Tsang,Cappellaro}, imaging~\cite{Kok,Giovannetti,app11146414,Tsang} and communication~\cite{BB84,Ekert,AcinPRL2007}, that can be accomplished in ways that are 
beyond what is possible classically. 
However, quantum information is notoriously fragile. As decoherence sets the boundary between quantum and classical physics, noise can easily wash away any potential quantum advantage \cite{bremner2017achieving,Noh2020efficientclassical,aharonov2023polynomial}.
Quantum error correction \cite{RevModPhys.87.307, roffe2019quantum,devitt2013quantum} has been developed as a tool to combat noise in quantum computing, as well as in sensing and communication, yet it requires large-scale quantum processing with substantial resource overhead~\cite{campbell2017roads}. 
Despite the rapid progress of recent years, we are still at a stage where quantum devices have limited control on a relatively small number of qubits~\cite{preskill2018quantum}.
In this context, techniques of error mitigation are being developed, see e.g.~Ref.~\cite{Bultrini2023unifying,qemit} and references therein. Though not necessarily scalable, these techniques aim at mitigating the harmful effects of noise on quantum information.

This work develops around the concept of error filtration, a scheme proposed for quantum communication~\cite{PhysRevA.72.012338,PhysRevLett.94.230501}
and computing \cite{vijayan2020robust,lee2022error} (see also~\cite{miguel2023sqem,miguel2023enhancing}).
The scheme exploits constructive and destructive interference to filter out dephasing noise with the help of ancillary vacuum modes.
Unlike other error mitigation schemes that are based on sampling or extrapolation, the approach considered here is hardware-based. 
Given a noisy quantum channel $\mathcal{E}$ acting on a optical mode, error filtration is implemented by calling $N$ i.i.d.~instances of $\mathcal{E}$, which are applied in parallel to the input mode and $N-1$ auxiliary vacuum modes. To harness inference, the $N$ modes are put in superposition by letting them pass through a multi-mode interferometer.


\begin{figure}[t!]
\centering
\includegraphics[width=0.9\linewidth]{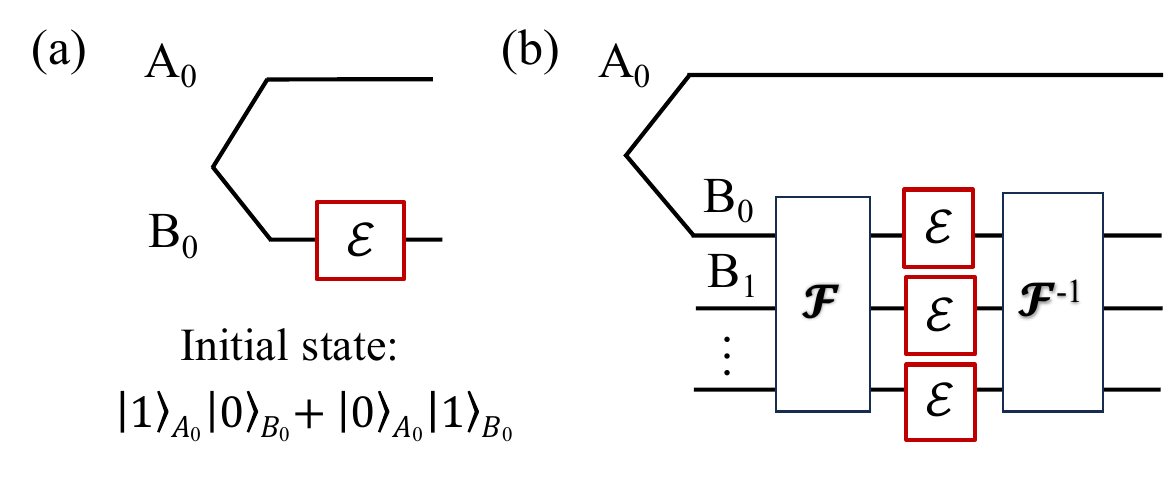}
\caption{Distribution of quantum information as encoded in a single-photon state in super-position over two optical modes, denoted as $A_0$, $B_0$.
In this model dephasing noise affects the $B$ modes: 
(a) no error filtration is used, as the quantum state is input into the noisy channel directly;
(b) error filtration is applied to mitigate noise on mode $B_0$, using $N-1$ vacuum ancillary modes, and a pair of $N$-port interferometers. For example, the interferometer may implement the Fourier transform and its inverse transformation.
}
\label{fig:scheme1}
\end{figure}


Here we are particularly interested in applications to stellar interferometry. 
One of the main challenges for distributing coherence and for realising a long-baseline interferometer is phase stabilisation, as thermal fluctuations lead to uncertainty in the path length of a fibre \cite{lim2021long}, which in turn degrade the quality of the signal.
We show that, by using linear optics and auxiliary vacuum modes, we can improve the task of parameter estimation. We use the quantum Fisher information as a figure of merit. 
 Unlike other works ~\cite{PhysRevA.72.012338,PhysRevLett.94.230501,miguel2023sqem,miguel2023enhancing,postmetro}, 
the two main applications we propose--coherent state distribution and stellar interferometry--achieve enhancements without requiring postselection.
Therefore, error mitigation in parameter estimation is deterministic and the quantum Fisher information is enhanced for all noise parameters. For clarity, we note that while this feature holds for our later schemes, the first illustrative scheme based on a single-photon state does rely on post-selection.
\color{black}
We investigate dephasing noise~\cite{lami2023exact}, but the scheme may be effective also against other noise models. However, it does not enhance robustness to loss. The latter may be achieved by nesting error filtration with schemes robust to photon loss by design, e.g.~Ref.~\cite{PhysRevLett.130.160801}.

The structure of the paper is as follows.
In Sec.~\ref{sec:noisem} we introduce the noise model and its range of applications.
We review the principle of error filtration in Sec.~\ref{sec:filter}, with its application in preserving quantum coherence in single-photon states.
The interferometer required by error filtration may introduce additional noise and loss; these effects are discussed in Sec.~\ref{sec:lossnoise}.
In Sec.~\ref{sec:coh} we discuss application in distributing coherent states along dephasing channels.
In Sec.~\ref{sec:interferometry} we discuss an application to boost the resolution of large-baseline telescopes without using quantum repeaters or quantum error correction~\cite{PhysRevLett.109.070503, PhysRevLett.123.070504, PhysRevA.100.022316, PhysRevLett.129.210502}. 
%
%
Further technical details are given in the Appendix~\ref{app:inter}, where we simulate imperfection in the beam splitters and show that with a 2\% variation on the reflectivities, the detrimental effect is minimal.
Also, we discuss the effects of detector dark counts.

\section{Noise model}\label{sec:noisem}

We assess the efficacy of error filtration to mitigate dephasing errors. Therefore, we consider a model of Bosonic dephasing channel~\cite{lami2023exact}, formally represented by the map $\mathcal{E}$ such as  
\begin{align}\label{defmap}
\ket{\psi}\rightarrow \mathcal{E}(\ket{\psi}\bra{\psi}) = 
\int p_\theta \, 
U_\theta | \psi \rangle \langle  \psi |  U_\theta^\dag  \, d\theta
\, ,
\end{align}
for some probability density distribution $p_\theta$ for the phase-shift angle $\theta$, where
\begin{align}
U_\theta  = e^{ i\hat{n}\theta}, \quad \int p_\theta \, d\theta  =1 \, ,
\end{align}
and $\hat n$ is the photon number operator.
We define the average dephasing parameter $\kappa$ as
\begin{align}
\kappa := \int  p_{\theta} e^{i \theta} \, 
d\theta
\, .
\end{align}

We focus on dephasing errors, but also other errors can be mitigated through error filtration, e.g.~depolarizing errors and generic Pauli errors~\cite{miguel2023sqem,miguel2023enhancing}. 
However, as discussed in Section~\ref{sec:lossnoise} the scheme is not suitable for mitigating loss.


Our application in stellar interferometry, discussed in Section~\ref{sec:interferometry}, may find application in arrays of telescope, e.g.~CHARA and the Very Large Telescope Interferometer.
Currently, dephasing noise is mainly affecting the measurement state (due to vibrations of the optical table) rather than the communication lines~\cite{McAlister_2005,Anugu_2020}.
However, future larger-scale arrays of telescopes may come with new challenges. For example, a future quantum internet may be used as a global infrastructure to make distant laboratories interfere in the optical domain~\cite{doi:10.1126/science.aam9288}. 
In general,  communication lines extending on longer distances, and not being single-purpose, may be affected by a variety of errors and in particular non-negligible dephasing noise.

\section{Error filtration on noisy single-photon states}\label{sec:filter}

In this Section we review the physical principle of error filtration~\cite{PhysRevA.72.012338}, applying it to the preservation of quantum coherence in single-photon states subject to dephasing noise.

Consider the task of distributing a state of a single photon in superposition over two optical modes with a fixed relative phase. 
States of this form play an important role in dual-rail logic~\cite{RevModPhys.79.135}, non-locality~\cite{PhysRevLett.66.252} and stellar interferometry~\cite{PhysRevLett.109.070503}.
Let $\{ a_{A_0}^\dag , a_{A_0} \}$ and $\{ a_{B_0}^\dag , a_{B_0}\}$ be the canonical Bosonic creation and annihilation operators for the two modes. 
Consider the state
\begin{align}
|\psi\rangle = 
\frac{1}{\sqrt{2}} (a_{A_0}^\dag + a_{B_0}^\dag) |0\rangle
=
\frac{1}{\sqrt{2}} \left( |10\rangle + |01\rangle \right)
\, ,
\end{align}
where $|0\rangle$ is the vacuum, $|10\rangle \equiv |1\rangle_{A_0} |0\rangle_{B_0}$ represents the state of a photon in mode $A_0$, and $|01\rangle \equiv |0\rangle_{A_0} |1\rangle_{B_0}$ that of a photon in mode $B_0$.
We want to distribute such a state to a distant laboratory by transmitting the mode $B_0$ through a noisy channel that induces dephasing. 
This is represented by the mapping in Eq.~(\ref{defmap}), where 
\begin{align}
U_\theta |\psi\rangle 
= \frac{1}{\sqrt{2}} \left( |10\rangle +  e^{i\theta} |01\rangle \right) \, .
\end{align}
We can readily compute the fidelity between the initial state and the final state at the output of the noisy channel,
\begin{align}\label{F1}
F_1 = \langle \psi | \mathcal{E}( |\psi\rangle \langle \psi |) 
| \psi\rangle 
& =
\frac{1}{2} \left( 1 + \mathsf{Re}\{\kappa\} \right)
\, .
\end{align}

\begin{figure}[t!]
\centering
\includegraphics[width=1.0\columnwidth]{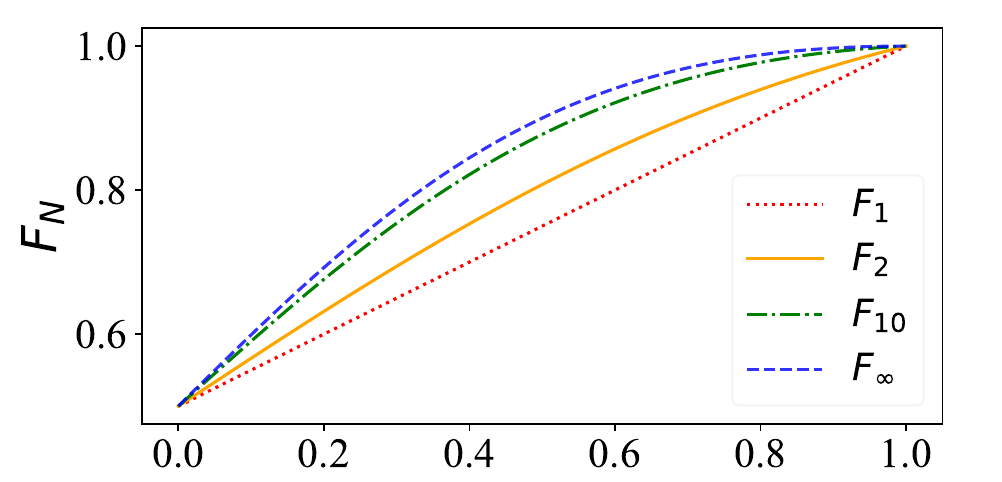}
\includegraphics[width=1.0\columnwidth]{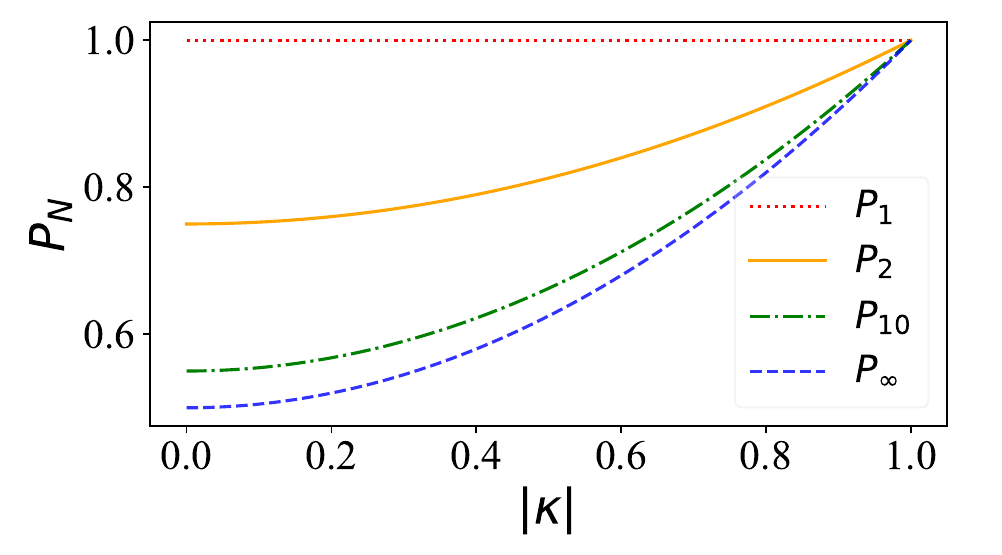}
\caption{Figures of merit for distribution of quantum information, when only the $B$ modes are subject to dephasing noise (as shown in Fig.~\ref{fig:scheme1}), assuming $\mathsf{Im}\{\kappa\} = 0$.
Top panel: fidelity $F_N$ plotted versus the noise parameter $|\kappa|$, computed from Eq.~(\ref{FN_1}); we show $N=1,2,10$, and in the limit that $N\to\infty$.
Bottom panel: post-selection probability $P_N$ versus the noise parameter $|\kappa|$, computed from Eq.~(\ref{PN_1}); from top to bottom for $N=1,2,10$, and in the limit that $N\to\infty$.
}
\label{fig:coherence_1}
\end{figure}

We show that with error filtration, i.e.~by exploiting vacuum ancillary modes and linear optics, we achieve a partial cancellation of the noise, resulting in increased fidelity.
To gain some intuition, we illustrate an example by using a single ancillary vacuum mode.
The ancillary mode, denoted $B_1$ is initially prepared in the vacuum state, then mixed with mode $B_0$ at a $50/50$ beam splitter.
The action of the beam splitter on the state of a single photon over two modes is represented by a Hadamard gate, yielding
\begin{align}
    |\psi'\rangle = \frac{1}{\sqrt{2}} \left( |100\rangle + \frac{1}{\sqrt{2}}|010\rangle + \frac{1}{\sqrt{2}}|001\rangle \right) \, ,
\end{align}
where $|lmn\rangle \equiv | l \rangle_{A_0} | m \rangle_{B_0} | n \rangle_{B_1}$ represents the state of a single-photon over three modes.
We assume that the modes $B_0$ and $B_1$ are affected by i.i.d.~dephasing noise; for example, the two modes may travel through identical but uncoupled optical fibers. 
For each value of the local phase-shift angles $\theta_1$, $\theta_2$, we have
\begin{align}
    U_{\theta_1} \otimes U_{\theta_2} |\psi'\rangle 
    =
    \frac{1}{\sqrt{2}} \left( |100\rangle + \frac{e^{i\theta_1}}{\sqrt{2}} |010\rangle  + \frac{e^{i\theta_2}}{\sqrt{2}} |001\rangle  \right)
    \, .
\end{align}

Finally, the modes are interfered again at a $50/50$ beam splitter, which accounts for applying a second Hadamard gate.
Post-selecting on measuring the vacuum state on the auxiliary mode $B_1$, we obtain the following (un-normalised) state on signal modes $A_0 B_0$
\begin{align}\label{phidopo}
    |\psi_{\theta_1\theta_2}\rangle 
    & =
    \frac{1}{\sqrt{2}} \left( |10\rangle 
    + \frac{e^{i\theta_1} + e^{i\theta_2}}{2} |01\rangle  
    \right) \, .
\end{align}
Note that within a single-photon model, post-selection on measuring the vacuum on the ancillary modes means considering only the events when the photon outputs in the signal mode $B_0$.

It remains to check that this state has fidelity larger than $F_1$, hence noise is effectively reduced.
We can anticipate that this is the case by inspection of Eq.~(\ref{phidopo}), where the noise factor appears to be the average of two independent random variables, therefore its variance is half the variance of $e^{i\theta_1}$ and $e^{i\theta_2}$.
By averaging over the random phase shifts we obtain the (un-normalised) state
\begin{align} 
    \rho_2
    & = \int d\mu_{\theta_1} d\mu_{\theta_1} | \psi_{\theta_1\theta_2} \rangle \langle \psi_{\theta_1\theta_2} | \\
    & =
\frac{1}{2} \bigg(
|10\rangle \langle 10| 
+
\frac{
1 + |\kappa|^2
}{2}
|01 \rangle \langle 01 |  
+
\kappa |01\rangle \langle 10| + \text{h.c.}
\bigg) \, .
\end{align}

The probability that the photon arrives at mode $B_0$ is
\begin{align}
    P_2 & = \mathrm{Tr}\rho_2 
    = \frac{1}{2} \left(
    1 + \frac{1+|\kappa|^2}{2} \right) \, ,
\end{align}
from which we finally obtain the fidelity upon normalisation
\begin{align}
F_2 = \frac{ \langle \psi | \rho_2 |\psi \rangle }{P_2} 
= 
\frac{1}{2} 
\left(
1 
+ \frac{ 
  4 \, \mathsf{Re}\{\kappa\}
 }
 {
3  + |\kappa|^2 } 
\right) \, ,
\end{align}
which is strictly greater than $F_1$ in Eq.~\eqref{F1} for any $|\kappa| \in (0,1)$.

We can extend the protocol to $N-1$ ancillary modes, $B_1$, $B_2$, $\dots$, $B_{N-1}$.
The ancillae are still prepared in the vacuum state, but the $50/50$ beam splitter is replaced by a $N$-mode interferometer. For example, we can choose one that implements the Fourier transform, with matrix elements
\begin{align}
\mathcal{F}_{hk} & = \frac{1}{\sqrt{N}} \, e^{i 2\pi h k / N} \, ,
\end{align}
for $h,k=0,1,\dots N-1$.
After the action of $N$ i.i.d.~dephasing channels, we apply the inverse Fourier transform,
\begin{align}
[ \mathcal{F}^\dag ]_{hk} & = \frac{1}{\sqrt{N}} \, e^{-i 2\pi h k / N} \, .
\end{align}
This is shown in Fig.~\ref{fig:scheme1}.

The final state, post-selected on the event that no photon arrive on the auxiliary modes, is described by the un-normalised state
\begin{align}\label{psimit}
|\psi_{\theta_1 \cdots \theta_N}\rangle
= \frac{1}{\sqrt{2}} \left( |10\rangle + |01\rangle 
\sum_{k=0}^{N-1} \frac{e^{i \theta_k}}{N}  \right) \, .
\end{align}

Note that here we have the average of $N$ i.i.d.~random variables, therefore we expect an $N$-fold reduction in the noise variance.
Averaging over noise realisations we obtain the post-selected state, which reads
\begin{align}
\rho_N 
& =
\frac{1}{2} 
\bigg(
|10\rangle \langle 10| +
\frac{
1 + (N-1) |\kappa|^2
}{N}
|01 \rangle \langle 01 |  
\nonumber \\
& \phantom{========}~
+
\kappa |01\rangle \langle 10| + \text{h.c.} 
\bigg) \, ,
\end{align}
from which we obtain the post-selection probability
\begin{align}\label{PN_1}
P_N = 
\frac{1}{2} \left( 1 
+ 
 \frac{
 1 + (N-1) |\kappa|^2
 }{N} 
\right) 
\end{align}
and the fidelity
\begin{align}\label{FN_1}
F_N 
= \frac{1}{2} \left( 
1 +
\frac{ 
 2 N \mathsf{Re}\{\kappa\}   
}{
1 + N 
+ (N-1) |\kappa|^2
} 
\right) \, .
\end{align}

Note that the fidelity is monotonically increasing with $N$, while the post-selection probability decreases. Eventually they approach limiting values
\begin{align}
F_\infty 
& = \frac{1}{2} \left( 
1 +
\frac{ 
 2 \mathsf{Re}\{\kappa\}
}{
1 + |\kappa|^2 }
\right)
\, , \\
P_\infty 
& = \frac{1}{2} \left( 
1 + |\kappa|^2 
\right) \, .
\end{align}

The fidelity is improved compared to the un-encoded case in Eq.~\eqref{F1} for all values of the noise parameter $\kappa$. 
We plot $F_N$ and $P_N$ in Fig.~\ref{fig:coherence_1} as a function of $|\kappa|$ assuming $\mathsf{Im}\{\kappa\}=0$.
The post-selection probability here is always larger than $1/2$, since there is a $50\%$ probability that the photon stays on mode $A_0$, which is noiseless.
%


\begin{figure}[t!]
\centering
\includegraphics[width=0.9\linewidth]{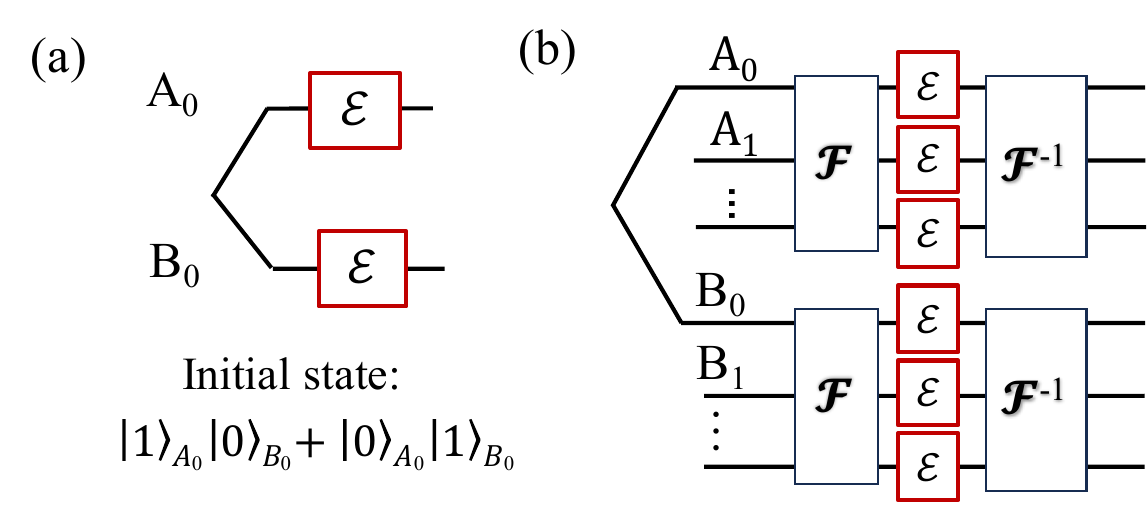}
\caption{Distribution of quantum information, as encoded in a single-photon state in super-position over two optical modes, $A_0$, $B_0$. 
In this setting, noise is applied independently to $A$ modes and $B$ modes: 
(a) no error filtration is applied, and the quantum state is input into the channel directly. (b) error filtration is applied to mitigate noise on both modes $A_0$ and $B_0$, using $2(N-1)$ vacuum ancillary modes, and four $N$-port interferometers implementing the Fourier transform and its inverse.}
\label{fig:ab_v2}
\end{figure}


\begin{figure}[t!]
\centering
\includegraphics[width=1\columnwidth]{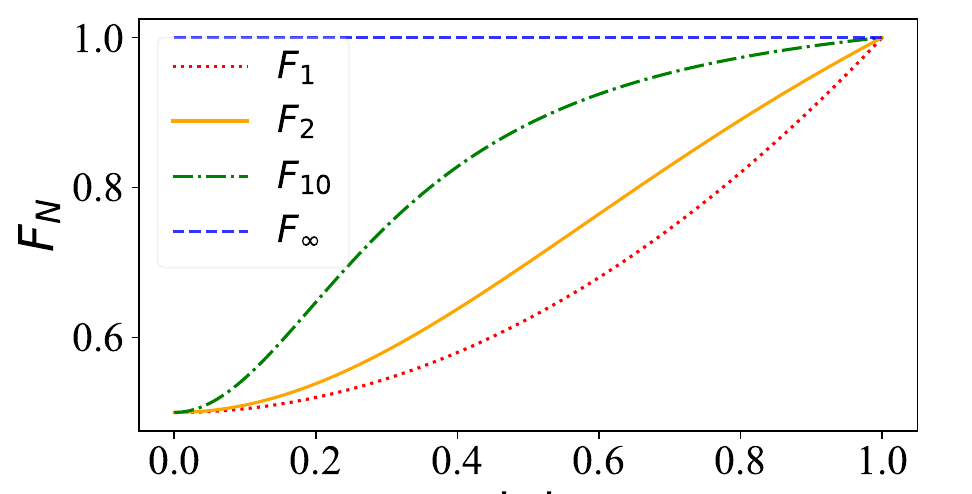}
\includegraphics[width=1\columnwidth]{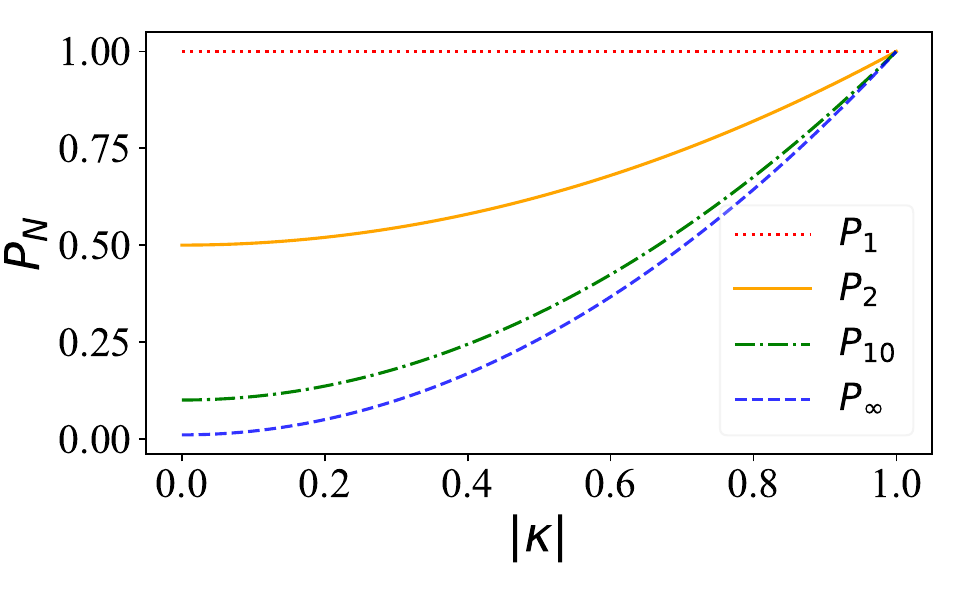}
\caption{Figures of merit of error filtration for distribution of quantum information in single-photons states, when both $A$ modes and $B$ modes are subject to dephasing noise.
Top panel: fidelity $F_N$ versus the noise parameter $|\kappa|$, computed from Eq.~(\ref{FN_2}), showing $N=1,2,10$, and in the limit that $N\to\infty$.
Bottom panel: post-selection probability $P_N$ is plotted versus the noise parameter $|\kappa|$, computed from Eq.~(\ref{PN_2}), showing $N=1,2,10$, and in the limit that $N\to\infty$.
For example, with $9$ ancillary modes we can improve the fidelity from $0.82$ to $0.97$ with a probability of success of about $68 \%$.
}
\label{fig:coherence_2}
\end{figure} 


If we assume that mode $A_0$ is also affected by i.i.d.~dephasing noise (as shown in Fig.~\ref{fig:ab_v2}), the post-selection probability may drop below $1/2$. However, in this symmetric case, if error filtration is applied to modes $A_0$ and $B_0$ independently, with the help of $2(N-1)$ vacuum modes, then the fidelity and post-selection probability read
\begin{align}\
F_N^\mathrm{sym} 
& = \frac{1}{2} \left( 1 + \frac{N \, |\kappa|^2}{ 1+ (N-1) |\kappa|^2} \right)
\, , \label{FN_2} \\
P_N^\mathrm{sym}
& = |\kappa|^2 + \frac{1 - |\kappa|^2}{N} 
\, .
\label{PN_2}
\end{align}

This shows that by increasing $N$ the fidelity approaches one, while the post-selection probability remains above $|\kappa|^2$.
These quantities are plotted in Fig.~\ref{fig:coherence_2}.
As an example, with $9$ ancillary modes we can improve the fidelity from $0.82$ to $0.97$ with a probability of success of about $68 \%$.



\section{Robustness to loss and noise in the interferometer}\label{sec:lossnoise}

While the error filtration scheme partially cancels dephasing noise, it is not able to mitigate loss. 
However, if there is only one photon travelling in the interferometer, the fidelity in Eq.~(\ref{FN_1}) still holds conditioned on the event that the photon is not lost.
If each branch of the interferometer has an equal attenuation factor $\eta$, then the state in Eq.~(\ref{psimit}) is replaced by
\begin{align}
    |\psi_{\eta;\theta_1 \cdots \theta_N}\rangle 
    = \frac{1}{\sqrt{2}} 
    \left( 
    |10 \rangle
    + \sqrt{\eta} 
    \sum_{k=0}^{N-1}
    \frac{e^{i\theta_k}}{N} 
        |01\rangle  
    \right) \, ,
\end{align}
and the fidelity reads
\begin{align}
F_N^\eta 
= \frac{1}{2} \left( 
1 +
\frac{ 
 2 N \sqrt{\eta} \, \mathsf{Re}\{\kappa\}   
}{
\eta + N 
+ \eta (N-1) |\kappa|^2
} 
\right) \, .
\end{align}

To achieve resilience to both dephasing and loss, error filtration may be nested with protocols that are robust to loss by design, e.g.~Ref.~\cite{PhysRevLett.130.160801}.

Most important, in practice introducing an $N$-mode interferometer will inevitably add loss and noise to the system.
To model additional dephasing, we assume that the dephasing noise in each branch of the interferometer is i.i.d.~and quantified by the parameter $\tilde\kappa > \kappa$, where in principle $\tilde\kappa$ may depend on $N$.
Also, we assume that the interferometer introduces loss with an overall attenuation factor $\eta$. 
Within this model, error filtration is still capable of enhancing fidelity as long as 
\begin{align}
\frac{ 
 2 N \sqrt{\eta} \, \mathsf{Re}\{ \tilde \kappa \} 
}{
\eta + N 
+ \eta (N-1) |\tilde \kappa|^2
} 
> \mathsf{Re}\{ \kappa \} \, .
\end{align}

\color{black}


\subsection{The even 1-to-$N$ beam splitter}
\label{sec:1-to-N}

Taking into account that any implementation of a multi-port interferometer brings additional noise and errors, we are interested in finding an interferometer design that has good error-filtration qualities without introducing too much noise.

A good choice is represented by a balanced beam splitter 
that transforms the first mode into an even superposition across the output.
In combination with its inverse, such an interferometer has exactly the same error-mitigating effect as the Fourier transform. 
The advantage of this scheme is that instead of needing $O(N^2)$ two-mode beam splitters \cite{PhysRevLett.73.58}, only $N-1$ two-mode beam splitters are needed for its implementation. 


\begin{figure}[h!]
\centering
\includegraphics[width=0.9\linewidth]{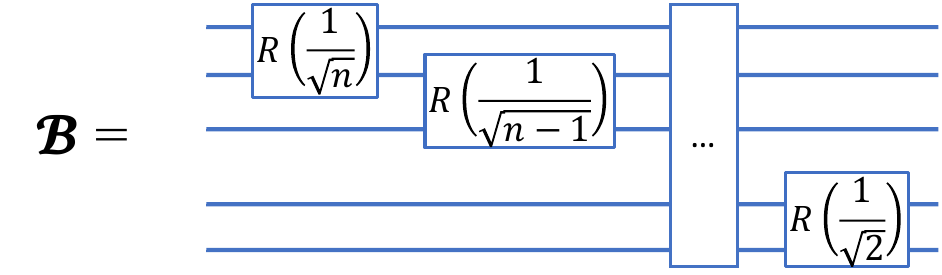}
\caption{ 
A balanced beam splitter that transforms the first mode into an even superposition, which can be decomposed into a series of 2-mode beam splitters; the reflectivity of the $i$th splitter is $r_i = 1/\sqrt{(N+1 -i)}$, $i \in [1,N-1]$.} 
\label{fig:even_splitter}  
\end{figure}


For example, in the case of four modes, the interferometer is represented by the matrix
\begin{align}
\mathcal{B}_4 =
\left(
\begin{array}{cccc}
 \frac{1}{2} & \frac{\sqrt{3}}{2} & 0 & 0 \\
 \frac{1}{2} & -\frac{1}{2 \sqrt{3}} & \sqrt{\frac{2}{3}} & 0 \\
 \frac{1}{2} & -\frac{1}{2 \sqrt{3}} & -\frac{1}{\sqrt{6}} & \frac{1}{\sqrt{2}} \\
 \frac{1}{2} & -\frac{1}{2 \sqrt{3}} & -\frac{1}{\sqrt{6}} & -\frac{1}{\sqrt{2}} 
\end{array}
\right) \, .
\end{align}
It is easy to check that this transformation yields the same output as the quantum Fourier transform when only the first input mode is populated. In general, for the case of $N$ modes, it is sufficient that the first column of unitary matrix has entries $1/\sqrt{N}$.



\color{black}

\section{Distribution of coherent states}
\label{sec:coh}
 
\begin{figure}[t]
\centering
\includegraphics[width=0.9\columnwidth]{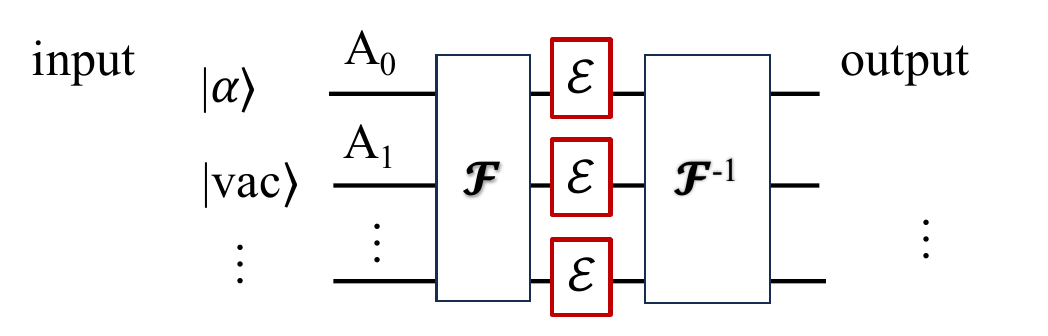}
\caption{
Scheme of error mitigation for distributing a coherent state through a dephasing channel. 
A coherent state $|\alpha\rangle$ is prepared for the input mode $A_0$.
Our encoding involves $N-1$ auxiliary modes, $A_1$, $\dots$, $A_{N-1}$ initially prepared in the vacuum state.
Signal and auxiliary modes are mixed by applying a balanced $N-$ splitter (an $N$-port interferometer that implements the quantum Fourier transform $\mathcal{F}$) at the input. Each mode is affected by i.i.d.~Bosonic dephasing channels $\mathcal{E}^{\otimes n}$. At the output, decoding is implemented by the interferometer that applies the inverse Fourier transform $\mathcal{F}^{-1}$.
Error mitigation is observed in the signal mode at the output.} 
\label{fig:coh_dist}  
\end{figure}

The mechanism of error filtration is not limited to single-photon states~\cite{PhysRevA.72.012338,PhysRevLett.94.230501}, and indeed the scheme also works in the framework of classical and semi-classical optics. Here we discuss its application to mitigate dephasing errors in coherent states. 
The latter are routinely used for sensing \cite{ip2022using}, phase locking \cite{hsieh1996phase}, classical and quantum \cite{minder2019experimental,grosshans2003quantum,wang2022twin} communications. Our techniques can be used to further improve the performance of such tasks. 
A schematic of our approach is depicted in Fig.~\ref{fig:coh_dist}.
An important difference with the case of single-photon states, is that error filtration does not require post-selection when applied to coherent states. 
This corresponds to having higher fidelity at the cost of reduced intensity.


Consider a coherent state that undergoes the Bosonic dephasing channel which is input in mode $A_0$. 
Applying a random phase $\theta$ to the state, we have
\begin{align}
\ket{\alpha} \rightarrow \ket{\alpha e^{i \theta}}_{A_0}.
\end{align}
From the expression for the overlap between two coherent states, $|\braket{\nu_2|\nu_1}|^2 = e^{-|\nu_1 - \nu_2|^2}$~\cite{mandel1995optical}, we obtain
\begin{align}
|\braket{\alpha|\alpha e^{i \theta_1}}|^2 = e^{- 2 |\alpha|^2 (1-\cos\theta)} \, ,
\end{align}
from which we compute the fidelity upon averaging over the noise realisations:\begin{align}
F^{(1)} = \int p_{\theta} e^{- 2 |\alpha|^2 (1-\cos\theta)} \, d\theta \, .
\end{align}
Using vacuum ancillary modes and linear optics we can achieve a partial cancellation of the noise, resulting in increased fidelity.  
First, consider the setup with one ancillary vacuum mode ($N=2$).
The ancillary mode, denoted $A_1$ is initially prepared in the vacuum state, then mixed with mode $A_0$ at a $50/50$ beam splitter. The state is acted upon by random phase shifts $\theta_1,\theta_2$:
\begin{align}
\biggl | \frac{\alpha}{\sqrt2}\biggr \rangle_{A_0} \biggl | \frac{\alpha}{\sqrt2}\biggr\rangle_{A_1}
\to 
\biggl | \frac{\alpha e^{i \theta_1} }{\sqrt2}\Biggr \rangle_{A_0} 
\biggl | \frac{\alpha e^{i \theta_2} }{\sqrt2}\Biggr \rangle_{A_1}.
\end{align}
We apply the second beam splitter:
\begin{align}
\biggl |\frac{\alpha e^{i \theta_1} }{\sqrt2}\biggr \rangle _{A_0}
\biggl |\frac{\alpha e^{i \theta_2} }{\sqrt2}\biggr \rangle_{A_1}
\to
\Biggl | \alpha \frac{ e^{i \theta_1}+e^{i \theta_2} }{2}\biggr \rangle _{A_0}
\Biggl | \alpha \frac{ e^{i \theta_1}-e^{i \theta_2} }{2}\biggr \rangle _{A_1}
\end{align}
Tracing over the second mode, and computing the overlap for specific values of $\theta_1,\theta_2$, yields
\begin{align}\label{eq:two_coherent}
\left| \left\langle \alpha  ~\bigg|\alpha \frac{ e^{i \theta_1}+e^{i \theta_2} }{2} \right\rangle \right|^2
&= e^{- \left|\alpha - \alpha \left(\frac{e^{i\theta_1} + e^{i\theta_2}}{2}\right) \right| ^2   },
\end{align}
Finally, the fidelity is obtained by taking the average over the noise realization,
\begin{align}
F^{(2)} 
&=\int p_{\theta_1}p_{\theta_2} \left| \left\langle \alpha~ \bigg| \alpha \frac{ e^{i \theta_1}+e^{i \theta_2} }{2} \right\rangle \right|^2 d\theta_1 d\theta_2  \, ,
\end{align}
which we expect to be higher than $F^{(1)}$, due to the fact the effective dephasing factor is the averaged of two statistically independent sources of dephasing.


\begin{figure}[t]
\centering
\includegraphics[width=1.0\columnwidth]{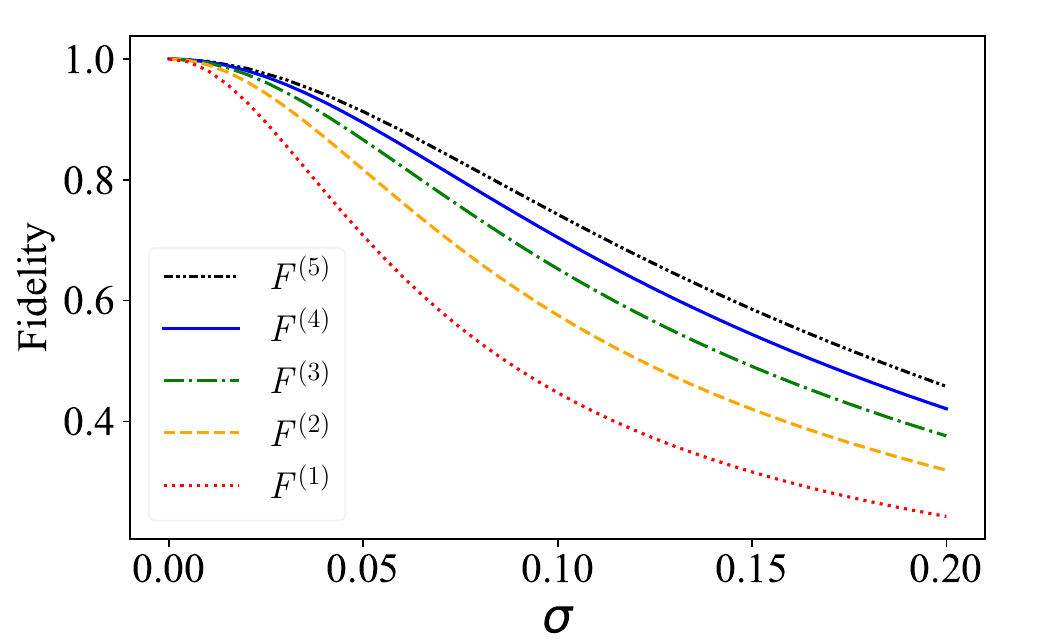}
\caption{
The figure shows the coherent state fidelity attained by error mitigation, plotted versus the noise parameter $\sigma$, as computed using Eq.~(\ref{eq:Fn_body}).
Dephasing errors are assumed to be Gaussian as in Eq.~(\ref{eq:gauss}), and the resulting fidelity is computed for different number of modes $N=1,2,3,4,5$ and mean photon number $|\alpha|^2 = 200$.
} 
\label{fig:coh_fid}  
\end{figure}

We can extend the protocol to $N-1$ ancillary modes, $A_1, A_2,...A_{N-1}.$ The ancillae are still prepared in the vacuum state, but the $50/50$ beam splitter is replaced by a $N$-mode interferometer. For example, we can choose one that implements the Fourier transform and its inverse, or the interferometer discussed in Section \ref{sec:1-to-N} that is equivalent but with lower complexity. 


When using $N-1$ ancillary channels, the quantity in Eq.~\eqref{eq:two_coherent} generalises to
\begin{align}\label{eq:n_modes_body}
 &\left|  \left\langle \alpha~ \bigg| \alpha \frac{\left(\sum_{k=1}^N e^{i\theta_k}\right) }{N} \right \rangle \right|^2
= e^{- \left|\alpha - \alpha \frac{\left (\sum_{k=1}^N e^{i\theta_k} \right)}{N}\right| ^2   } \, , 
\end{align}
and the attained fidelity is
\begin{align} \label{eq:Fn_body}
F^{(N)} = \int d{\bm \theta} \prod_{i=1}^N p_{\theta_i} \exp\left[- \left|\alpha - \alpha \frac{\left(\sum_{k=1}^N e^{i\theta_k} \right)}{N}\right| ^2        \right] \, ,
\end{align}
where ${\bm \theta} = [\theta_1,\theta_2,...\theta_N]$
and
$d{\bm \theta} = d\theta_1 d\theta_2 \cdots d\theta_N$.

Note that here we have the average of $N$ i.i.d.~random variables in the first term, therefore we expect an $N$-fold reduction in the noise variance. 
We plot Eq.~(\ref{eq:Fn_body}), for $N = 1,2,3,4$ and $N=5$ in Fig.~\ref{fig:coh_fid}, computed under the assumption that each random phase shift is Gaussian-distributed with variance $\sigma^2 \ll 1$ centered at $0$,
\begin{align}
\label{eq:gauss}
p_\theta = \frac{1}{\sqrt{2\pi} \sigma}e^{-\frac{1}{2}\left(\theta/\sigma \right)^2 },
\quad \sigma \ll 1.
\end{align}
For higher $N$ the calculation becomes intensive due to the multi-variable integration, though we see that that a large improvement of fidelity can be gained by using only a few modes.

\section{Stellar interferometry}
\label{sec:interferometry}

\begin{figure}[t!]
\centering
\includegraphics[width=1.0\columnwidth]{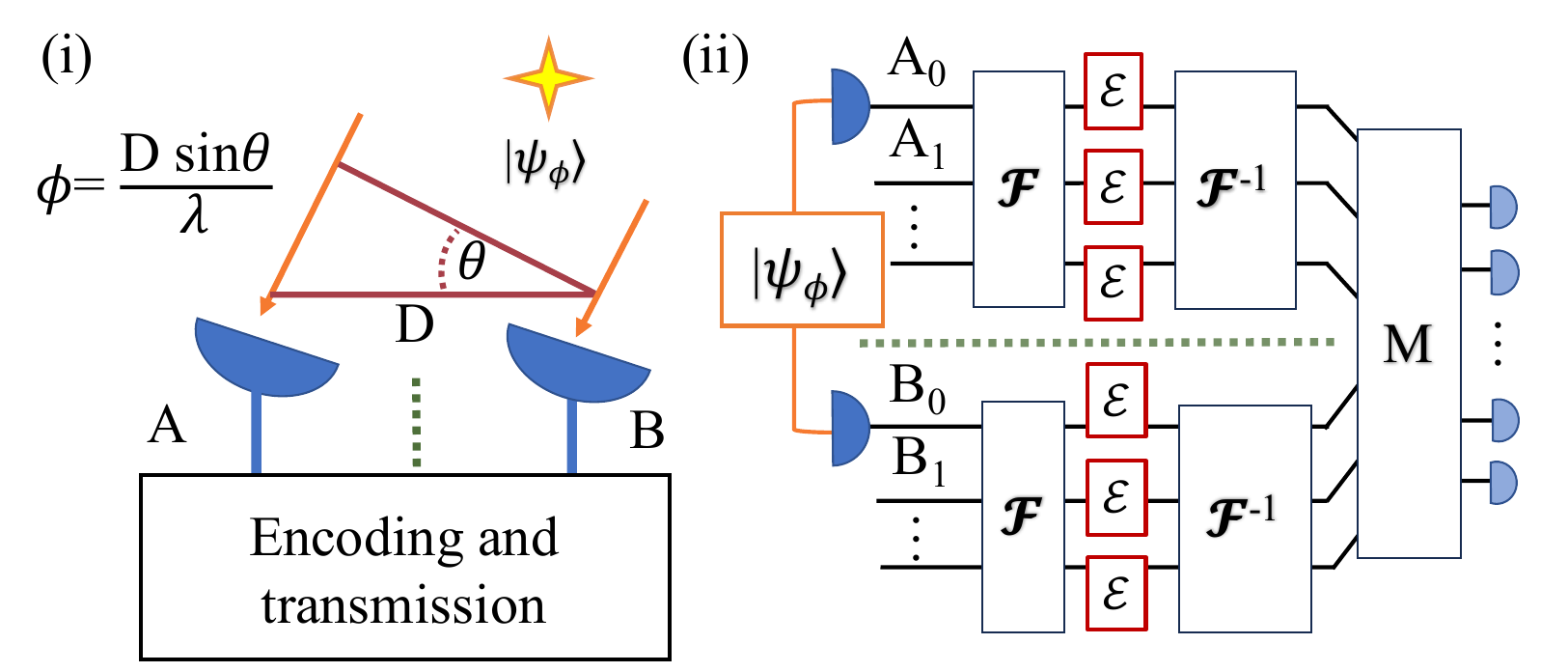}
\caption{\label{fig:scheme} 
Application of the error mitigation scheme to stellar interferometry. 
(i) Light from a distant source is collected at two sites $A$ and $B$ that are spatially separated; the goal is to estimate the angle 
$\theta$, given that the transmission lines from the collection points to the measurement station are affected by dephasing noise.
(ii) Details of the error mitigation scheme. To mitigate phase noise, ancillary vacuum modes $A_1$, $\dots$, $A_{N-1}$ and $B_1$, $\dots$, $B_{N-1}$ are introduced locally. 
Before transmission, modes $A$ and $B$ are encoded using a $N$-port interferometer implementing the Fourier transform $\mathcal{F}$ (or an equivalent interferometer), after the transmission, the inverse transform is applied before the measurement $M$.}
\end{figure}

In the context of stellar interferometry, consider a two-site scenario, telescope station $A$ and telescope station $B$; for quantum-enhanced telescopy \cite{PhysRevLett.109.070503}, $A$ and $B$ are separated by large distances. This setup is depicted in Fig.~\ref{fig:scheme}. One of the main challenges in astronomical imaging is the the requirement for phase stabilisation across the interferometer, which our scheme can help improve.

In the weak-signal limit, we model the stellar state as a single photon that is received by either of the two telescopes. 
Such a state may be formally described as \cite{PhysRevLett.129.210502, PhysRevLett.107.270402, PhysRevLett.109.070503, PhysRevLett.123.070504}
\begin{align}
\hat \rho_\phi = \left(\frac{1+\gamma}{2}\right) \ket{\psi_\phi^+ }\bra{\psi_\phi^+} + 
            \left(\frac{1-\gamma}{2}\right) \ket{\psi_\phi^-}\bra{\psi_\phi^-} \, ,
\end{align}
where
\begin{align}
\ket{\psi_\phi^\pm} = \frac{1}{\sqrt2}\left(\ket{01}_{A_0 B_0} \pm  e^{i\phi}\ket{10}_{A_0 B_0}\right) \, .
\end{align}
The parameter $\phi \in [0 ,2\pi )$ is related to the location of the sources 
and $\gamma \in [0,1]$ is related to the Fourier transform of the source intensity distribution
(the shape of the objects) via the van Cittert--Zernike theorem~\cite{mandel1995optical}. If $\gamma=1$, the object is a single point, and $\gamma$ decreases as the size of the object increases. \color{black}
Here we focus on the problem of finding the ultimate precision limit in the estimation 
of the parameters $\phi$ and $\gamma$.

The ultimate precision in parameter estimation is specified by the quantum Cram\'er--Rao bound~\cite{caves,caves1} (see also \cite{giovannetti2011advances,giovannetti2006quantum}).
For estimation of a parameter $\varphi$ encoded into a quantum state
$ \hat \rho_\varphi =\sum_i \lambda_i \ket{i}\bra{i}$, 
Cram\'er--Rao sets a lower bound on the variance $(\Delta \varphi)^2 = \langle \varphi^2 \rangle - \langle \varphi \rangle^2$ of any unbiased estimator~$\varphi$:
\begin{align} \label{eq:var}
 (\Delta  \varphi) ^2 \geqslant  \frac{1}{n  J_\varphi( \hat \rho_\varphi)} \, ,
\end{align}
where $n$ is the number of copies of $\hat \rho_\varphi$ and $J_\varphi$ is the quantum Fisher information (QFI) associated with the state $\hat \rho_\varphi$,
\begin{align}
J_\varphi(\hat \rho_\varphi) = \sum_{i,j; \lambda_i +\lambda_j \neq 0} 2\frac{|\braket{i|\partial_\varphi \hat \rho_\varphi|j}|^2}{\lambda_i+\lambda_j}.
\end{align}
%
If there are multiple parameters we want to estimate, where $\vec\varphi = (\varphi_1, \varphi_2, \dots)$, we can define a QFI matrix $\mathbf{J}$
that quantifies not only the QFI for each parameter (diagonal components) but also for correlated parameters (off-diagonal components).
The matrix elements are given by
\begin{align}
J_{jk}
:=  \frac{1}{2}\text{Tr}[\hat \rho_{\vec{\varphi}}(\hat L_j \hat L_k + \hat L_k \hat L_j)],
\end{align}
 where $\hat L_j$ is the symmetric logarithmic derivative with respect to $\varphi_j$~\cite{paris2009quantum}. 

The inverse of the QFI matrix provides a lower bound on the covariance matrix
 $\big[ \text{Cov}(\vec\varphi) \big]_{jk} = \braket{\theta_j \varphi_k } - \braket{\varphi_j}\braket{\varphi_k}$,
 \begin{align}
\text{Cov}\left(\vec\varphi \right) \geq \frac{1}{N } \vec J^{-1}.
\end{align}
For a single parameter, the Cram\'er--Rao bound is known to be attainable \cite{barndorff2000fisher}.
For multiple parameters, the bound is not always attainable because the optimal measurement operators that saturate the bound for the individual parameters may not commute. Therefore, the parameters may not be simultaneously measurable.

We refer to Refs.~\cite{Paris,Sidhu} for detailed discussions of the QFI and its properties.
In this case, the parameters $\gamma$ and $\phi$ are not compatible because their SLD's do not commute, an hence they and cannot be estimated optimally simultaneously ~\cite{Pearce2017optimal,PhysRevA.109.052434,PhysRevLett.129.210502}. In the worst-case scenario, we can estimate $\phi$ and $\gamma$ in separate measurements, which means the covariance matrix would be effectively increased by a factor two.
\color{black}

As local measurements would not allow us to measure $\phi$ or $\gamma$, we need to send light collected by the distant telescopes towards a common measurement station, where the two paths are interfered. We assume that the measurement station is in the middle between the two telescopes. 
En route from the telescopes to the measurement station, modes $A_0$ and $B_0$ are affected by i.i.d.~dephasing noise, which can be modelled as the Bosonic dephasing channel introduced above.

Suppose we send the signal through the channel directly without encoding, the output density matrix is
\begin{align} \label{eq:no_encoding}
&\mathcal{E}_{A_0} \circ \mathcal{E}_{B_0}(\hat \rho_\phi) = \nn
& \frac{(1+\gamma |\kappa|^2)}{2} \ket{\psi_\phi^+}\bra{\psi_\phi^+} +  \frac{(1-\gamma |\kappa|^2)}{2} \ket{\psi_\phi^-}\bra{\psi_\phi^-} \, .
\end{align}

For the state in Eq.~\eqref{eq:no_encoding}, we can calculate the QFI of the parameters $\phi$ and $\gamma$,
\begin{align}
\label{eq:no_enc_phi}
J_\phi^{(1)} &= \gamma^2 |\kappa|^4 \, , \qquad
J_\gamma^{(1)} = \frac{|\kappa| ^4}{1-\gamma ^2 \kappa ^4} \, .
\end{align}

We now show that error filtration yields up to a quadratic factor of improvement in the QFI in terms of $\kappa$.
Due to the operational meaning of the QFI, and for a fair comparison, we will consider the QFI of the average state without post-selection on the ancillary modes.

In the simplest setting, we introduce ancillary modes $A_1$ and $B_1$.
Hadarmard gates (i.e.~$50/50$ beam splitters) couple mode $A_0$ with $A_1$ and, independently, mode $B_0$ with $B_1$. Explicitly, for the state $\ket{\psi_\phi^+}$, this yields 
\begin{align}
\ket{\psi_\phi^+} \to 
\frac{1}{2} \left( 
\ket{10;00} + \ket{01;00} 
+ e^{i\phi} \ket{00;10} 
+ e^{i\phi} \ket{00;01} 
\right) \, ,
\end{align}
where $|lm;np\rangle \equiv |l\rangle_{A_0} |m\rangle_{A_1} |n\rangle_{B_0} |p\rangle_{B_1}$.
After passing through the dephasing channels, every off-diagonal component will pick up the factor $|\kappa|^2$, therefore the density matrix, expressed in the basis $\left\{ \ket{10;00} , \ket{01;00} , \ket{00;10} , \ket{00;01} \right\}$, reads 

\begin{align}
\mathcal{E}^{\otimes 4}(\hat \rho_\phi) =
\frac{1}{4}
\left(
\begin{array}{cccc}
 1 & |\kappa|^2 & \gamma e^{i \phi }|\kappa|^2 & \gamma  e^{i \phi }|\kappa|^2 \\
 |\kappa|^2 & 1 & e^{i \phi }|\kappa|^2 & e^{i \phi }|\kappa|^2 \\
 \gamma  e^{-i \phi }|\kappa|^2 & \gamma  e^{-i \phi }|\kappa|^2 & 1 & |\kappa|^2 \\
 \gamma  e^{-i \phi }|\kappa|^2 & \gamma  e^{-i \phi }|\kappa|^2 & |\kappa|^2 & 1 \\
\end{array}
\right) \, .
\end{align}

Finally, this state is passed through another pair of independent Hadamard gates, yielding
\begin{align}\label{QFIrho2}
\hat \rho_{\phi;2} =
%
\frac{1}{2}
\left(
\begin{array}{cccc}
 \frac{1+|\kappa|^2}{2} & 0 & e^{-i \phi } |\kappa|^2 & 0 \\
 0 & \frac{1-|\kappa|^2}{2} & 0 & 0 \\
 e^{i \phi } |\kappa|^2 & 0 & \frac{1+|\kappa|^2}{2} & 0 \\
 0 & 0 & 0 & \frac{1-|\kappa|^2}{2} 
\end{array}
\right) \, .
\end{align}
This shows that the photon appears in modes $A_0$ and $B_0$ with probability 
\begin{align}
 P_2 =\frac{1 + |\kappa|^2}{2} 
 \, ,
\end{align}
carrying information about the parameters $\phi$ and $\gamma$. 
If the photon ends in modes $A_1B_1$, it bears no information about these parameters. 
From this observation, we can compute the associated QFI, giving
\begin{align} \label{main:eq:two_modes}
J_\phi^{(2)}   &= \frac{2 \gamma ^2 \kappa ^4}{1+|\kappa| ^2} \, , \quad
J_\gamma^{(2)} = \frac{2 \left(\kappa ^6+\kappa ^4\right)}{\left(1-4 \gamma ^2\right) \kappa ^4+2 \kappa ^2+1} \, .
\end{align}


\begin{figure}[t!]
\centering
\includegraphics[width=1.0\columnwidth]{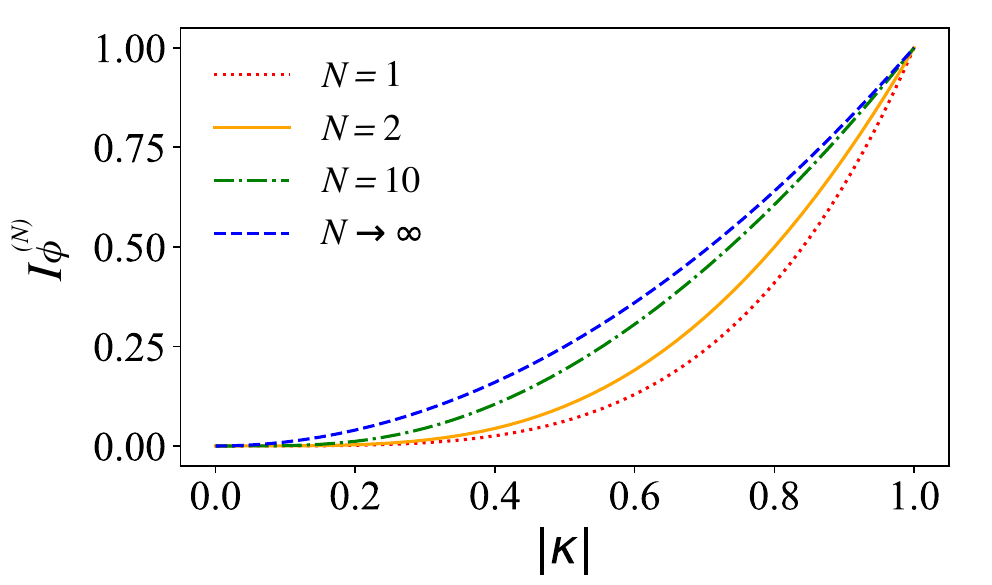}
\includegraphics[width=1.0\columnwidth]{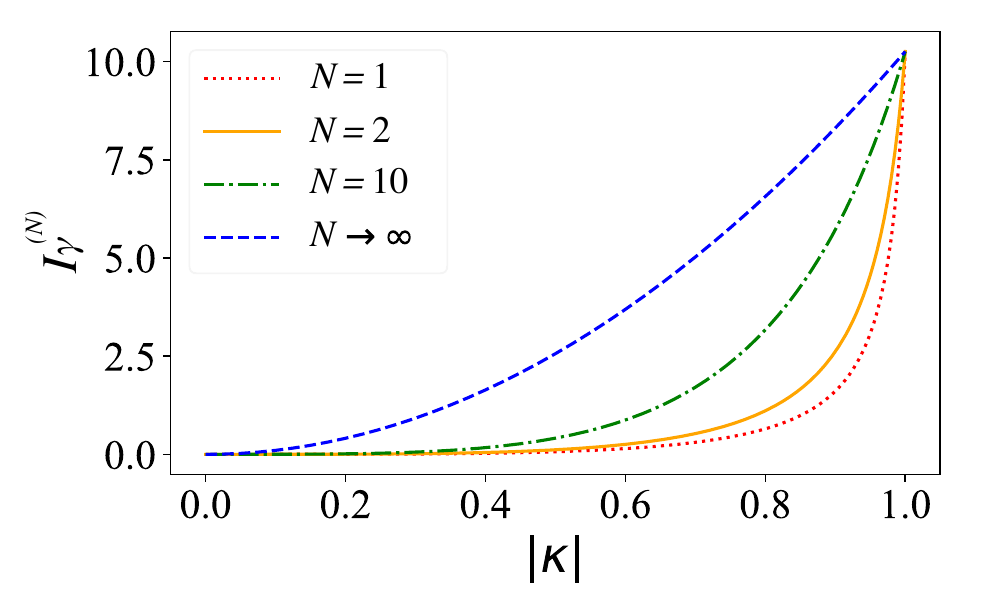}
\caption{
(Top) QFI for the estimation of the phase parameter $\phi$, plotted as a function of the noise parameter $|\kappa|$, as computed in Eq.~(\ref{IQ_N}). 
We show $N=1$ (red dotted line),  $N=2$ (orange solid line)  $N=10$ (green dotted dashed line) and in the limit that $N\to\infty$ (blue dashed line), {for $\gamma = 1$}.
(Bottom) the QFI for the coherence parameter $\gamma$, plotted as a function of $|\kappa|$ as in Eq.~(\ref{eq:gamma_n}); {here we choose to plot $\gamma = 0.95$. }
\label{fig:qfi_kappa} } 
\end{figure}


We can now generalise the protocol to using $2(N-1)$ ancillary modes, $A_1$, $\dots$, $A_N$ and $B_1$, $\dots$, $B_N$.
To each group of modes we apply error filtration.
After the application of the inverse Fourier transform, we find that, as for the case $N=2$,
only when the photon appears in modes $A_0 B_0$, 
an event that happens with probability 
\begin{align}
P_N = \frac{1}{N}(1+(N-1)|\kappa|^2) \, ,    
\end{align}
the photon carries information about the phase $\phi$ and the parameter $\gamma$.
For the estimation of the parameter $\phi$ we obtain
\begin{align}\label{IQ_N}
J_\phi^{(N)}= \frac{\gamma ^2 |\kappa| ^4 N}{|\kappa| ^2 (N-1)+1} \, , \\
\lim_{N\rightarrow \infty} 
J_\phi^{(N)} 
= \gamma^2|\kappa|^2 \, .
\end{align}
{ 
In the limit that ${N\rightarrow \infty}$, we gain a quadratic factor in $|\kappa|$. }
We plot the QFI for $N=1,2,10$ and $N\rightarrow \infty$ in 
Fig.~\ref{fig:qfi_kappa}. We see a notable gap for all values of $|\kappa|$, and the QFI monotonically increases by increasing the number of auxiliary modes. 
Similarly, for the estimation of the parameter $\gamma$, we obtain
\begin{align}\label{eq:gamma_n}
J_\gamma^{(N)}=  \frac{1}{2} |\kappa| ^4 N \bigg(&\frac{1}{\kappa ^2 (-\gamma  N+N-1)+1}+ \nn
&\frac{1}{\kappa ^2 (\gamma  N+N-1)+1}\bigg) \, , \\
\lim_{N\rightarrow \infty} 
J_\gamma^{(N)} = & \frac{\kappa ^2}{1-\gamma ^2} \, .
\end{align}
 The effect of noise is reduced by a factor larger than $|\kappa|^2$ compared to the un-encoded case. 
Here we have chosen to show the plots for $\gamma =0.95$. See Appendix~\ref{app:inter} for details of the calculation and additional plots for $\gamma = 0.5$ and $0.8$. 

In Appendix~\ref{app:inter} we also investigate the effects of detector dark counts and imperfect reflectivities of the beam splitters in implementing the encoding/decoding operations. Note that detector dark counts occurs independently, and does not affect our error mitigation scheme. For imperfect beam splitter reflectivities, we show that even having a standard deviation of 2\% on each consistuent component has minimal effect for most all values of $|\kappa|$.




\color{black}
\section{Discussions and Conclusions}

We have proposed to use interferometry to mitigate dephasing noise that afflicts photonic quantum information carriers.
We have discussed applications of the general scheme of error filtration to shield coherent states and to improve stellar interferometry.
Our approach uses only passive linear optics and vacuum ancillary modes, without the need for ancillary entanglement or optical non-linearity. 
In stellar interferometry, we show that the quantum Fisher information can be increased by up to a quadratic factor, hence improving phase estimation in the presence of dephasing noise.

The optimal performance of our scheme is achieved asymptotically for a large number of auxiliary modes, that is, by means of an interferometer with many input ports.
In practice, increasing complexity of the interferometer unavoidably comes with additional sources of noise and loss.
However, in most cases, a relatively low number of modes is sufficient to improve significantly the quality of the signal getting close to the asymptotic performance.
For example, our results suggest that a ten-mode interferometer already yields near-optimal results even for strong noise (see e.g.~Fig.~\ref{fig:qfi_kappa}). 
For weak noise, an even smaller-size interferometer may suffice.

For long-haul quantum communications, the state-of-the-art visibility parameter/error rate is between 2$\%$ \cite{PhysRevX.9.021046} to 7$\%$ \cite{amies2023quantum} (with phase locking).
Our scheme can be use in addition to existing phase-locking techniques. While a 98\% (corresponding to $|\kappa|=0.98$ visibility may be sufficient for quantum communication, the resolution of very small astronomical objects can benefit from having a visibiliy parameter above 99$\%$. In a recent proof-of-principle experimental collaboration by the same authors \cite{zanforlin2022optical}, we showed that an optical interferometer with 99.5$\%$ visibility is able to out-perform diffraction-limited direct imaging by 3-4 orders of magnitude.

\color{black}

Our error mitigation scheme is most naturally implemented on integrated photonic circuits. 
Given the recent experimental progress, where research groups globally have demonstrated chips that are reconfigurable~\cite{Notaros:17}, high-fidelity~\cite{politi2009integrated,moody20222022} and high-transmission \cite{PhysRevLett.123.250503}, we expect our scheme to become a viable way to mitigate dephasing noise, paving a pathway to the practical application of quantum technology.


\begin{acknowledgments}
This project has received funding from the: 
European Union’s Horizon Europe research and innovation programme under the project "Quantum Secure Networks Partnership" (QSNP, grant agreement No 101114043);
European Union, Next Generation EU: PNRR MUR project PE0000023-NQSTI;
and 
Italian Space Agency, project `Subdiffraction Quantum Imaging' (SQI) n.~2023-13-HH.0.
ZH thanks the generous hospitality of UTS:QSI during which this work was carried out. ZH is supported by a Sydney Quantum Academy Postdoctoral Fellowship
and an ARC DECRA Fellowship (DE230100144) ``Quantum-enabled super-resolution imaging".
\end{acknowledgments}

\appendix
\widetext


%

\section{Coherent state distribution}


\subsection{No encoding}

Now we calculate the fidelity of the coherent state at the output of the channel.
Apply a random phase $\theta_1$ to the state, we have
\begin{align}
\ket{\alpha} \rightarrow \ket{\alpha e^{i \theta_1}}_{A_0}
\end{align}
For two coherent states $\ket{\nu_1}$ and $\ket{\nu_2}$, the overlap is 
\begin{align}
|\braket{\nu_2|\nu_1}|^2 = e^{-|\nu_1 - \nu_2|^2}
\end{align}

Thus,
\begin{align}
|\braket{\alpha|\alpha e^{i \theta_1}}|^2 \nn
&= e^{-|\alpha - \alpha e^{i\theta_1}|^2} \nn
&= e^{-(\alpha - \alpha e^{i\theta_1})(\alpha^*-\alpha^*e^{-i\theta_1})} \nn
&= e^{-(|\alpha|^2 - |\alpha|^2 e^{-i\theta_1} - |\alpha|^2 e^{i\theta_1}+ |\alpha|^2 ) } \nn
&= e^{- 2 |\alpha|^2 (1-\cos\theta_1)}
\end{align}

To get the fidelity after the noise channel,  we integrate over $\theta_1$ 
\begin{align}
F^{(1)} = \int d\theta_1 ~ \mu_{\theta_1} e^{- 2 |\alpha|^2 (1-\cos\theta_1)}
\end{align}

\subsection{Two or more modes}

Now, to gain some intuition, we examine using two modes. We write the coherent state being split with a 50:50 beam splitter, with two  random phase-shifts $\theta_1,\theta_2$:
\begin{align}
\biggl | \frac{\alpha}{\sqrt2}\biggr \rangle_{A_0} \biggl | \frac{\alpha}{\sqrt2}\biggr\rangle_{A_1}
\to 
\biggl | \frac{\alpha e^{i \theta_1} }{\sqrt2}\Biggr \rangle_{A_0} 
\biggl | \frac{\alpha e^{i \theta_2} }{\sqrt2}\Biggr \rangle_{A_1}.
\end{align}

We apply the second beam splitter :
\begin{align}
\biggl |\frac{\alpha e^{i \theta_1} }{\sqrt2}\biggr \rangle _{A_0}
\biggl |\frac{\alpha e^{i \theta_2} }{\sqrt2}\biggr \rangle_{A_1}
\to
\Biggl | \alpha \frac{ e^{i \theta_1}+e^{i \theta_2} }{2}\biggr \rangle _{A_0}
\Biggl | \alpha \frac{ e^{i \theta_1}-e^{i \theta_2} }{2}\biggr \rangle _{A_1}
\end{align}

We trace over the second mode:
\begin{align}
\left| \alpha \frac{ e^{i \theta_1}+e^{i \theta_2} }{2}\right\rangle_{A_0}
\left| \alpha \frac{ e^{i \theta_1}-e^{i \theta_2} }{2}\right\rangle_{A_1}
\to 
\left| \alpha \frac{ e^{i \theta_1}+e^{i \theta_2} }{2}\right\rangle_{A_0}
\end{align}
Compute the overlap for specific values of $\theta_1,\theta_2$:
\begin{align}\label{ap:eq:two_coherent}
\left| \left\langle \alpha  ~\bigg|\alpha \frac{ e^{i \theta_1}+e^{i \theta_2} }{2} \right\rangle \right|^2
&= e^{- \left|\alpha - \alpha \left(\frac{e^{i\theta_1} + e^{i\theta_2}}{2}\right) \right| ^2   },
\end{align}
Take the expression inside the exponential
\begin{align}
&\left|\alpha - \alpha \left(\frac{e^{i\theta_1} + e^{i\theta_2}}{2}\right) \right| ^2 =
\left(\alpha -  \alpha \left(\frac{e^{i\theta_1} + e^{i\theta_2}}{2}\right)  \right) 
\left(\alpha^* -  \alpha^* \left(\frac{e^{-i\theta_1} + e^{-i\theta_2}}{2}\right)  \right) \nn
&= |\alpha|^2 - |\alpha|^2\left(\frac{e^{-i\theta_1} + e^{-i\theta_2}}{2}\right) 
              - |\alpha|^2\left(\frac{e^{i\theta_1} + e^{i\theta_2}}{2}\right)  
              +  |\alpha|^2  \left(\frac{e^{-i\theta_1} + e^{-i\theta_2}}{2}\right)
                          \left(\frac{e^{i\theta_1} + e^{i\theta_2}}{2}\right) \nn
&=\frac{1}{2} |\alpha| ^2 (3+\cos (\theta_1-\theta_2)-2 \cos (\theta_1)-2 \cos (\theta_2))
\end{align}

Therefore
\begin{align}
F^{(2)} 
&=\int \int d\theta_1 d\theta_2 ~ \mu_{\theta_1}\mu_{\theta_2} 
 \left| \left\langle \alpha ~\bigg| \alpha \frac{ e^{i \theta_1}+e^{i \theta_2} }{2} \right\rangle \right|^2 \nn
&=\int \int d\theta_1 d\theta_2 ~ \mu_{\theta_1}\mu_{\theta_2} 
 \exp\left[- \frac{1}{2} |\alpha| ^2 \left(3+\cos (\theta_1-\theta_2)-2 \cos (\theta_1)-2 \cos (\theta_2) \right)    \right]
\end{align}

When using $N-1$ ancillary channels, the quantity in Eq.~\eqref{ap:eq:two_coherent} generalises to

\begin{align}\label{eq:n_modes}
  \left|  \left\langle\alpha| \alpha  \left(\frac{\sum_{k=1}^N e^{i\theta_k} }{N}\right) \right\rangle  \right|^2
&= \exp\left[- \left|\alpha - \alpha \left(\frac{\sum_{k=1}^N e^{i\theta_k} }{N}\right) \right| ^2    \right] \nn
\end{align}
Finally, the fidelity for using $N-$ modes is
\begin{align} \label{eq:F_n}
F^{(N)} = \int d{\bm \theta} \prod_{i=1}^N \mu_{\theta_i} \exp\left[- \left|\alpha - \alpha \left(\frac{\sum_{k=1}^N e^{i\theta_k} }{N}\right) \right| ^2        \right]
\end{align}
where ${\bm \theta} = [\theta_1,\theta_2,...\theta_N]$.

\section{Direct interferometry -- detailed calculations}\label{app:inter}




In the weak-photon limit, a single photon is received by Alice and Bob at a time. The state can be written as
\begin{align}
\hat \rho_\phi = \left(\frac{1+\gamma}{2}\right) \ket{\psi_\phi^+ }\bra{\psi_\phi^+} + 
            \left(\frac{1-\gamma}{2}\right) \ket{\psi_\phi^-}\bra{\psi_\phi^-} 
\end{align}
where
\begin{align}
\ket{\psi_\phi^\pm} = \frac{1}{\sqrt2}\left(\ket{01}_{A_0 B_0} \pm  e^{i\phi}\ket{10}_{A_0 B_0}\right) .
\end{align}

\subsection{No encoding}
\noindent
Suppose we send the collected stellar signal through the channel directly without encoding, 
\begin{align}
\hat \rho_\phi \rightarrow \mathcal{E}_{A_0} \circ \mathcal{E}_{B_0}(\rho_\phi) 
\end{align}

Let us examine effect of the channel on the the off-diagonal components, e.g
\begin{align}
\mathcal{E}(\ket{10}\bra{01}) &=
\int d\theta d\theta' \mu_{B_1}(\theta)  \mu_{B_2}(\theta') e^{i \theta} e^{-i\theta'}\ket{10}\bra{01} \nn
&= |\kappa|^2\ket{10}\bra{01}
\end{align}

We now calculate this state explicitly 
\begin{align} \label{ap:eq:no_encoding}
\mathcal{E}_{A_0} \circ \mathcal{E}_{B_0}(\hat \rho_\phi) = 
&\frac{1+\gamma}{2}\left[ \frac{1}{2}\left( \ket{01}\bra{01}+  \ket{10}\bra{10} \right) +
 \left(e^{-i\phi} |\kappa|^2 \ket{01}\bra{10} + e^{i\phi} |\kappa|^2 \ket{10}\bra{01} \right) \right] + \nn
&
\frac{1-\gamma}{2}\left[ \frac{1}{2}\left( \ket{01}\bra{01}+  \ket{10}\bra{10} \right) -
 \left(e^{-i\phi} |\kappa|^2 \ket{01}\bra{10} + e^{i\phi} |\kappa|^2 \ket{10}\bra{01} \right) \right] \nn
&= \frac{(1+\gamma |\kappa|^2)}{2} \ket{\psi_\phi^+}\bra{\psi_\phi^+} +  \frac{(1-\gamma |\kappa|^2)}{2} \ket{\psi_\phi^-}\bra{\psi_\phi^-}.
\end{align}

\noindent

For the state in Eq.~\eqref{ap:eq:no_encoding}, we can calculate the QFI of the parameters:
\begin{align}
\label{ap:eq:no_enc_phi}
J_\phi^{(1)} = \gamma^2 |\kappa|^4, \qquad
J_\gamma^{(1)} = \frac{\kappa ^4}{1-\gamma ^2 \kappa ^4}.
\end{align}
Here the superscript $(1)$ denotes that 1 mode is used.

\subsection{Two-mode encoding}

Now let us look at the scenario where we use two modes, where the modes $A_0 (B_0)$ is coupled to the mode $A_1 (B_1)$ with the Hadamard gate
\begin{align} 
H = \frac{1}{\sqrt2} \left(
\begin{array}{cc}
 1 & 1 \\
 1 & -1 \\
\end{array}
\right)
\end{align}

\begin{align}
\hat a^\dag_{A_0} \rightarrow \frac{1}{\sqrt 2} ( \hat a^\dag_{A_0} +\hat a^\dag_{A_1}      )
\end{align}

\begin{align}
\ket{\psi_\phi^\pm}    &=  \frac{1}{\sqrt2}(\ket{10}_{A_0 B_0} +  e^{i\phi}\ket{01}_{A_0 B_0})\nn
        &\rightarrow    \frac{1}{2}(\ket{1000}_{A_0 A_1 B_0 B_1}+ \ket{0100}_{A_0 A_1 B_0 B_1} 
                       \pm   e^{i\phi}\ket{0010}_{A_0 A_1 B_0 B_1} \pm e^{i\phi}\ket{0001}_{A_0 A_1 B_0 B_1} )
\end{align}
\color{black}

Since this state has support on the single-photon Fock basis, we will define
\begin{align}
\ket{1000}_{A_0 A_1 B_0 B_1} \equiv \left(
\begin{array}{c}
 1 \\
 0 \\
 0 \\
 0 \\
\end{array}
\right), \quad
\ket{0100}_{A_0 A_1 B_0 B_1} \equiv \left(
\begin{array}{c}
 0 \\
 1 \\
 0 \\
 0 \\
\end{array}
\right)\nn
\ket{0010}_{A_0 A_1 B_0 B_1} \equiv \left(
\begin{array}{c}
 0 \\
 0 \\
 1 \\
 0 \\
\end{array}
\right), \quad
\ket{0001}_{A_0 A_1 B_0 B_1} \equiv \left(
\begin{array}{c}
 0 \\
 0 \\
 0 \\
 1 \\
\end{array}
\right)
\end{align}

The density matrix can be written as
\begin{align}
\hat \rho_\phi 
=\frac{1}{4}
\left(
\begin{array}{cccc}
 1 & 1 & \gamma e^{i \phi } & \gamma e^{i \phi } \\
 1 & 1 & \gamma e^{i \phi } & \gamma e^{i \phi } \\
 \gamma e^{-i \phi } & \gamma e^{-i \phi } & 1 & 1 \\
 \gamma e^{-i \phi } & \gamma e^{-i \phi } & 1 & 1 \\
\end{array}
\right)
\end{align}


Every off-diagonal component will pick up the same factor $|\kappa|^2$, therefore the density matrix becomes
\begin{align}
\hat \rho'_\phi =\mathcal{E}^{\otimes 4}(\rho_\phi) =
\frac{1}{4}
\left(
\begin{array}{cccc}
 1 & |\kappa|^2 & \gamma e^{i \phi }|\kappa|^2 & \gamma  e^{i \phi }|\kappa|^2 \\
 |\kappa|^2 & 1 & e^{i \phi }|\kappa|^2 & e^{i \phi }|\kappa|^2 \\
 \gamma  e^{-i \phi }|\kappa|^2 & \gamma  e^{-i \phi }|\kappa|^2 & 1 & |\kappa|^2 \\
 \gamma  e^{-i \phi }|\kappa|^2 & \gamma  e^{-i \phi }|\kappa|^2 & |\kappa|^2 & 1 \\
\end{array}
\right)
\end{align}

We can apply $\mathcal{F}^{-1}$, which is simply two sets of 50:50 BS for two modes
\begin{align}
F_2^{-1} = \frac{1}{\sqrt2}
\left(
\begin{array}{cccc}
 1 & 1 & 0 & 0 \\
 1 & -1 & 0 & 0 \\
 0 & 0 & 1 & 1 \\
 0 & 0 & 1 & -1 \\
\end{array}
\right)
\end{align}


The output is
\begin{align}
{\hat \rho''}_\phi = F_2^{-1} {\rho'}_\phi {F_2^{-1}}^\dag =
\left(
\begin{array}{cccc}
 \frac{1}{4} \left(1+|\kappa|^2\right) & 0 & \frac{1}{2} \gamma |\kappa|^2 e^{i \phi } & 0 \\
 0 & \frac{1}{4} \left(1-|\kappa|^2\right) & 0 & 0 \\
 \frac{1}{2} |\kappa|^2 \gamma e^{-i \phi } & 0 & \frac{1}{4} \left(1+|\kappa|^2\right) & 0 \\
 0 & 0 & 0 & \frac{1}{4} \left(1-|\kappa|^2\right) \\
\end{array}
\right)
\end{align}

\noindent
We can see that the modes $\ket{0100}$ and $\ket{0001}$ are uncoupled, and these have 0 eigenvalues if $\kappa=1$. The eigenvalues and vectors of $\rho''_\phi$ are

\begin{align}
\frac{1}{4} \left(1-|\kappa|^2\right), &\quad  [0,0,0,1]^T \nn
\frac{1}{4} \left(1-|\kappa|^2\right), &\quad  [0,1,0,1]^T \nn
\frac{1}{4} \left( 1+ (1-2 \gamma ) |\kappa |^2\right), &\quad \frac{1}{\sqrt2} \left[ 1,0,-e^{-i \phi },0\right]^T \nn
\frac{1}{4} \left(1+ (1+2 \gamma )  |\kappa| ^2\right),&\quad  \frac{1}{\sqrt2} \left[1,0,e^{-i \phi },0\right]^T
\end{align}

We can compute the QFI, giving
\begin{align} \label{ap:eq:two_modes}
I_\phi^{(2)}   &= \frac{2 \gamma ^2 \kappa ^4}{1+|\kappa| ^2}, \qquad
I_\gamma^{(2)} = \frac{2 \left(\kappa ^6+\kappa ^4\right)}{\left(1-4 \gamma ^2\right) \kappa ^4+2 \kappa ^2+1}.
\end{align}




\subsection{N modes}

Using linearity, if the input to the channel is

\begin{align}
\label{eq:lin}
\hat \rho_\phi = \left(\frac{1+\gamma}{2} \right) \ket{\psi_\phi^+}\bra{\psi_\phi^+} +
       \left(\frac{1-\gamma}{2} \right) \ket{\psi_\phi^-}\bra{\psi_\phi^-}
\end{align}

Then the output will be
\begin{align}
\mathcal{E}^{\otimes 2N} (\hat  \rho_\phi) = 
\left(\frac{1+\gamma}{2} \right) \mathcal{E}^{\otimes 2N}\left(\ket{\psi_\phi^+}\bra{\psi_\phi^+}\right) +
\left(\frac{1-\gamma}{2} \right)  \mathcal{E}^{\otimes 2N}\left(\ket{\psi_\phi^-}\bra{\psi_\phi^-}\right)
\end{align}

First, let us analyse the $ \ket{\psi_\phi^+}\bra{\psi_\phi^+}$ component. After the encoding and the dephasing channel, the density matrix will take the form

\begin{align}
\hat \rho'_\phi = 
\left(
\begin{array}{cc}
 C & D \\
 D^* & C \\
\end{array}
\right)
\end{align}
where
\begin{align}
C \equiv\frac{1}{2N}
\left(
\begin{array}{ccc}
 1 & |\kappa|^2 & |\kappa|^2\\
 |\kappa|^2 & ... & ... \\
 |\kappa|^2 & ... &  1 \\
\end{array}
\right),
\qquad
D\equiv 
\frac{e^{i\phi} |\kappa|^2}{2N}\left(
\begin{array}{ccc}
 1 & ... & 1\\
 ... & ... & ... \\
 1 & ... &  1 \\
\end{array}
\right), \qquad 
D^* \equiv 
\frac{e^{-i\phi} |\kappa|^2}{2N}\left(
\begin{array}{ccc}
 1 & ... & 1\\
 ... & ... & ... \\
 1 & ... &  1 \\
\end{array}
\right)
\end{align}

The inverse QFT between the two sets of modes is a direct sum of two N-mode inverse QFT's:

\begin{align}
\mathcal{Q}^{-1} = \left(
\begin{array}{cc}
 \mathcal{F}^{-1} & 0 \\
 0 &  \mathcal{F}^{-1} \\
\end{array}
\right)
\end{align}

\begin{align}
\hat  \rho''_\phi &= \mathcal{Q}^{-1}\rho'_\phi \mathcal{Q}   \nn
            &= 
\left(
\begin{array}{cc}
 \mathcal{F}^{-1} & 0 \\
 0 & \mathcal{F}^{-1} \\
\end{array}
\right)
\left(
\begin{array}{cc}
 C & D \\
 D^* & C \\
\end{array}
\right)
\left(
\begin{array}{cc}
 \mathcal{F} & 0 \\
 0 & \mathcal{F} \\
\end{array}
\right)
\nn
&=
\left(
\begin{array}{cc}
 \mathcal{F}^{-1} C \mathcal{F} & \mathcal{F}^{-1} D \mathcal{F} \\
 \mathcal{F}^{-1} D^* \mathcal{F} &  \mathcal{F}^{-1} C \mathcal{F} \\
\end{array}
\right)
\end{align}

\color{black}

Since $D$ and $D^*$ are uniform, performing an inverse QFT gives a single component
\begin{align}
{\mathcal{F}^{-1}} D \mathcal{F} = \frac{e^{i\phi} |\kappa|^2}{2}
\left(
\begin{array}{ccc}
 1 & 0 & 0 \\
 0 & 0 & 0\\
 0 & 0  & ...
\end{array}
\right), \qquad
{\mathcal{F}^{-1}} D^* {\mathcal{F}} = \frac{e^{-i\phi} |\kappa|^2}{2}
\left(
\begin{array}{ccc}
 1 & 0 & 0 \\
 0 & 0 & 0\\
 0 & 0  & ...
\end{array}
\right)
\end{align}

Since only the coherent term between $C$ and $D$ contain information on $\phi$, we only need to calculate 
$[\mathcal{F}^{-1} C {\mathcal{F}}]_{1,1}$. These matrices are highly symmetric, which simplifies the calculation:
\begin{align}
{\mathcal{F}}^{-1} C {\mathcal{F}} &= 
\frac{1}{\sqrt N}
\left(
\begin{array}{ccc c}
 1 & 1               & 1   & 1    \\
 1 & e^{-i 2\pi/N }   & ... & ...  \\
 1 & ... &  ...      & ...   \\
 1 & ... & ...    & ...  
\end{array}
\right)
 \times
\frac{1}{2N}
\left(
\begin{array}{ccc c}
 1 & |\kappa|^2 & |\kappa|^2 & ...\\
 |\kappa|^2 & 1 & |\kappa|^2  &...  \\
 |\kappa|^2 & |\kappa|^2 &  1      & ...\\
 ... &     ...  &   ... & ...  \\
\end{array}
\right) 
\times  
\frac{1}{\sqrt N}
\left(
\begin{array}{ccc c}
 1 & 1               & 1   & 1    \\
 1 & e^{i 2\pi/N }   & ... & ...  \\
 1 & ... &  ...      & ...   \\
 1 & ... & ...    & ...  
\end{array}
\right) \nn
{\mathcal{F}}^{-1} C {\mathcal{F}} &=
\frac{1}{ N}
\left(
\begin{array}{ccc}
 1 & 1 & 1\\
 1 & e^{-i 2\pi/N }   & ... \\
 1 & ... &  .. \\
\end{array}
\right)
\frac{1}{2N}
\left(
\begin{array}{ccc}
 1+(N-1)|\kappa|^2 &  ... & ...\\
 1+(N-1)|\kappa|^2& ... & ... \\
 1+(N-1)|\kappa|^2 & ... &  ... \\
\end{array}
\right) 
\end{align}
Therefore,
\begin{align}
{\mathcal{F}}^{-1} C {\mathcal{F}}_{1,1} &=\frac{1}{2N}(1+ (N-1)|\kappa|^2 )
\end{align}

This means that we can examine only a $2\time 2$ sub matrix of the density matrix to calculate the QFI. The relevant components are:
\begin{align}
\hat  \rho_\phi^{+ '} &=
\left(
\begin{array}{cc}
 \left[{\mathcal{F}}^{-1} C {\mathcal{F}}\right]_{1,1}   & \left[{\mathcal{F}}^{-1} D {\mathcal{F}}\right]_{1,1}       \\
  \left[\mathcal{F}^{-1} D^* {\mathcal{F}}\right]_{1,1}  &   [\mathcal{F}^{-1} C {\mathcal{F}}]_{1,1} \\
\end{array}
\right)\nn
&= \frac{1}{2}
\left(
\begin{array}{cc}
 \frac{1}{N}(1+ (N-1)|\kappa|^2 ) & e^{i\phi}|\kappa|^2  \\
 e^{-i\phi}|\kappa|^2 & \frac{1}{N}(1+ (N-1)|\kappa|^2 \\
\end{array}
\right)
,
\end{align}
noting this is not normalised.

Similarly
\begin{align}
\hat  \rho_\phi^{-'} =  \frac{1}{2}
\left(
\begin{array}{cc}
 \frac{1}{N}(1+ (N-1)|\kappa|^2 ) & -e^{i\phi}|\kappa|^2  \\
 -e^{-i\phi}|\kappa|^2 & \frac{1}{N}(1+ (N-1)|\kappa|^2 \\
\end{array}
\right)
\end{align}

Therefore overall, the components we need to consider to calculate the QFI are
\begin{align}
\left(\frac{1+\gamma}{2} \right) \hat  \rho_\phi^{+ '} +  \left(\frac{1-\gamma}{2} \right) \hat  \rho_\phi^{- '} 
= \left(\begin{array}{cc}
 \frac{1}{N}(1+ (N-1)|\kappa|^2 ) & \gamma e^{i\phi}|\kappa|^2  \\
 \gamma e^{-i\phi}|\kappa|^2 & \frac{1}{N}(1+ (N-1)|\kappa|^2 \\
\end{array}
\right)
\label{eq:final_rho_prior_detection}
\end{align}

The two eigenvalues and vectors that are sensitive to $\phi$ are
\begin{align}
e_1 &=\frac{1+\kappa ^2 (-\gamma  N+N-1)}{2 N},\quad v_1 = \left[1,- e^{-i\phi}\right]^T/\sqrt2\nn
e_2 &=\frac{1+\kappa ^2 (\gamma   N+N-1)}{2 N} ,\quad v_2 = \left[1, e^{-i\phi} \right]^T/\sqrt2
\end{align}
And the QFI is
\begin{align}
I_\phi^{(N)} = \frac{\gamma ^2 \kappa ^4 N}{\kappa ^2 (N-1)+1} \nn
\lim_{N\rightarrow \infty} \left[I_\phi^{(N)}  \right] = \gamma^2|\kappa|^2
\end{align}
We see that in the limit of large $N$, the QFI is boosted by a quadratic factor in $|\kappa|$ compared to the unencoded case.

For $\gamma$
\begin{align}
I_\gamma^{(N)} &= \frac{1}{2} \kappa ^4 N \left(\frac{1}{\kappa ^2 (-\gamma  N+N-1)+1}+\frac{1}{\kappa ^2 (\gamma  N+N-1)+1}\right) \nn
\lim_{N\rightarrow \infty} &\left[I_\gamma^{(N)} \right] = \frac{\kappa ^2}{1-\gamma ^2}
\end{align}
The effect of noise is significantly reduced compared to the unencoded case.

\begin{figure}
\centering
\includegraphics[width=0.45\columnwidth]{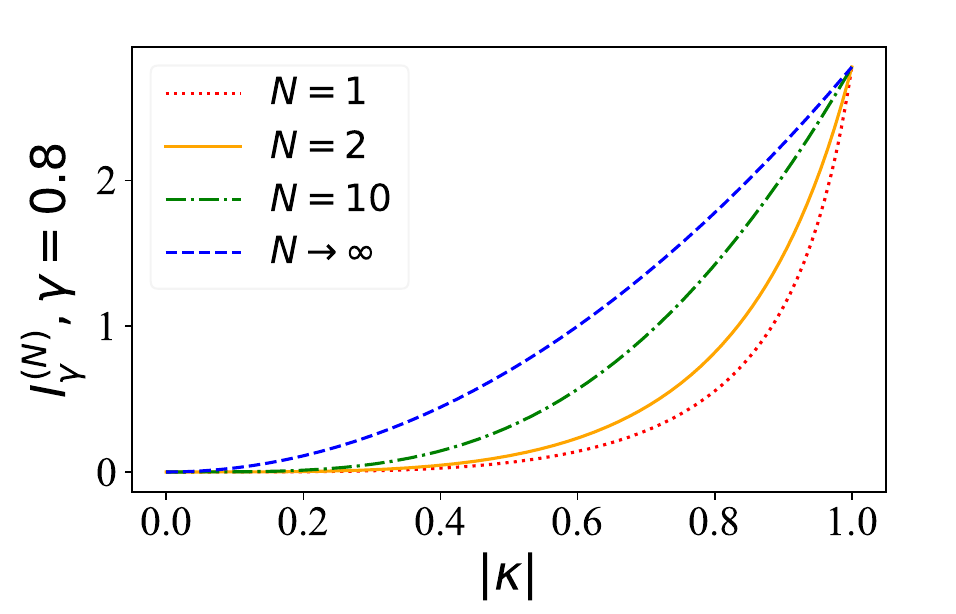}
\includegraphics[width=0.45\columnwidth]{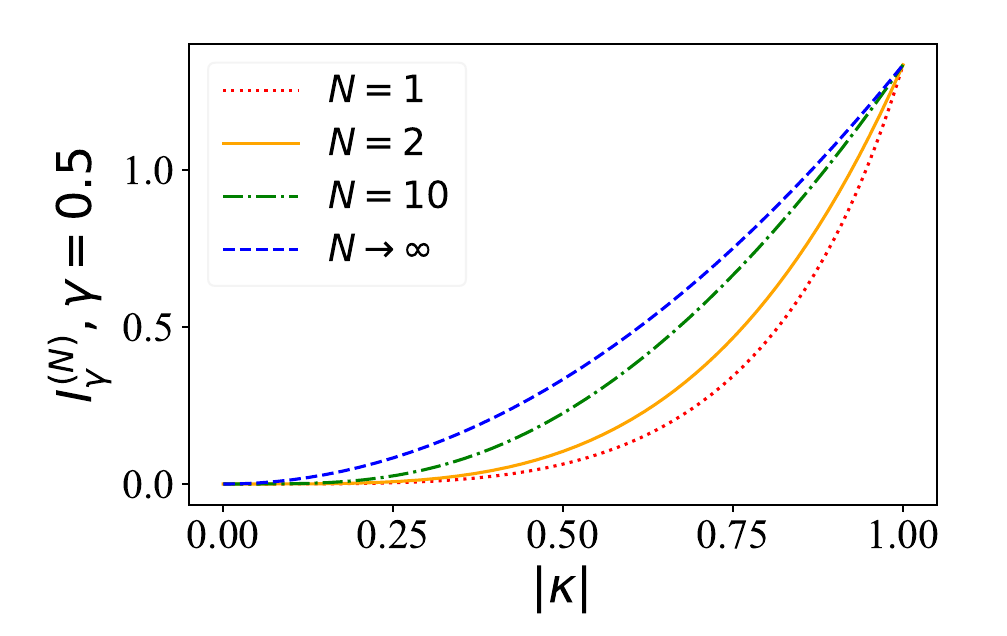}
\caption{The QFI for $\gamma = 0.5, 0.8$.} 
\end{figure}


\subsection{Beam splitter inefficiencies}
\label{sec:uneven_bs}
For the 4-mode splitters
\begin{align}
\mathcal{B}_4 =
\left(
\begin{array}{cccc}
 \frac{1}{2} & \frac{\sqrt{3}}{2} & 0 & 0 \\
 \frac{1}{2} & -\frac{1}{2 \sqrt{3}} & \sqrt{\frac{2}{3}} & 0 \\
 \frac{1}{2} & -\frac{1}{2 \sqrt{3}} & -\frac{1}{\sqrt{6}} & \frac{1}{\sqrt{2}} \\
 \frac{1}{2} & -\frac{1}{2 \sqrt{3}} & -\frac{1}{\sqrt{6}} & -\frac{1}{\sqrt{2}} \\
\end{array}
\right),\qquad
\mathcal{B}_4^{-1} = \left(
\begin{array}{cccc}
 \frac{1}{2} & \frac{1}{2} & \frac{1}{2} & \frac{1}{2} \\
 \frac{\sqrt{3}}{2} & -\frac{1}{2 \sqrt{3}} & -\frac{1}{2 \sqrt{3}} & -\frac{1}{2 \sqrt{3}} \\
 0 & \sqrt{\frac{2}{3}} & -\frac{1}{\sqrt{6}} & -\frac{1}{\sqrt{6}} \\
 0 & 0 & \frac{1}{\sqrt{2}} & -\frac{1}{\sqrt{2}} \\
\end{array}
\right) \nonumber
\end{align}
we simulate beam splitter inefficiencies/fabrication error by adding i.i.d Gaussian noise to the reflectivities. More explicitly, $\mathcal{B}_4$ is can be implemented by a the followin series of two-mode splitters:

\begin{align}
R_1 = 
\left(
\begin{array}{cccc}
 r_1 & \sqrt{1-r_1^2} & 0 & 0 \\
\sqrt{1-r_1^2} & -r_1 & 0 & 0 \\
 0 & 0 & 1 & 0 \\
 0 & 0 & 0 & 1 \\
\end{array}
\right),\quad
R_2 = 
\left(
\begin{array}{cccc}
 1 & 0 & 0 & 0 \\
 0 & r_2 & \sqrt{1-r_2^2} & 0 \\
 0 & \sqrt{1-r_2^2} & -r_2 & 0 \\
 0 & 0 & 0 & 1 \\
\end{array}
\right), \quad
R_3 = 
\left(
\begin{array}{cccc}
 1 & 0 & 0 & 0 \\
 0 & 1 & 0 & 0 \\
 0 & 0 & r_3 & \sqrt{1-r_3^2} \\
 0 & 0 & \sqrt{1-r_3^2} & -r_3 \\
\end{array}
\right)
\end{align}

\begin{figure}[h!]
\centering
\includegraphics[width=0.45\columnwidth]{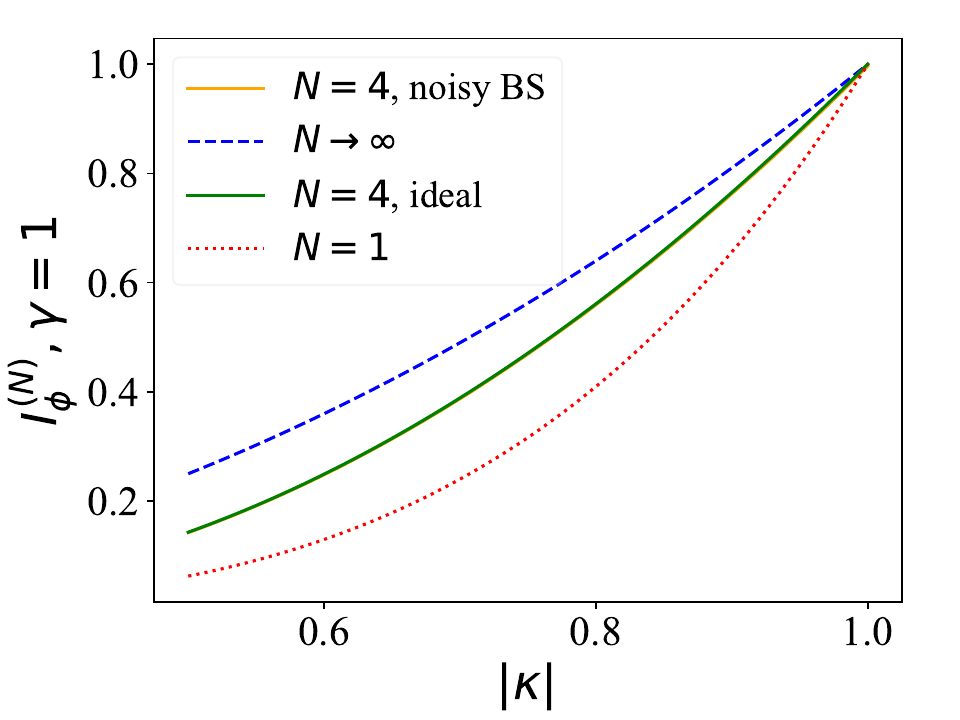}
\includegraphics[width=0.45\columnwidth]{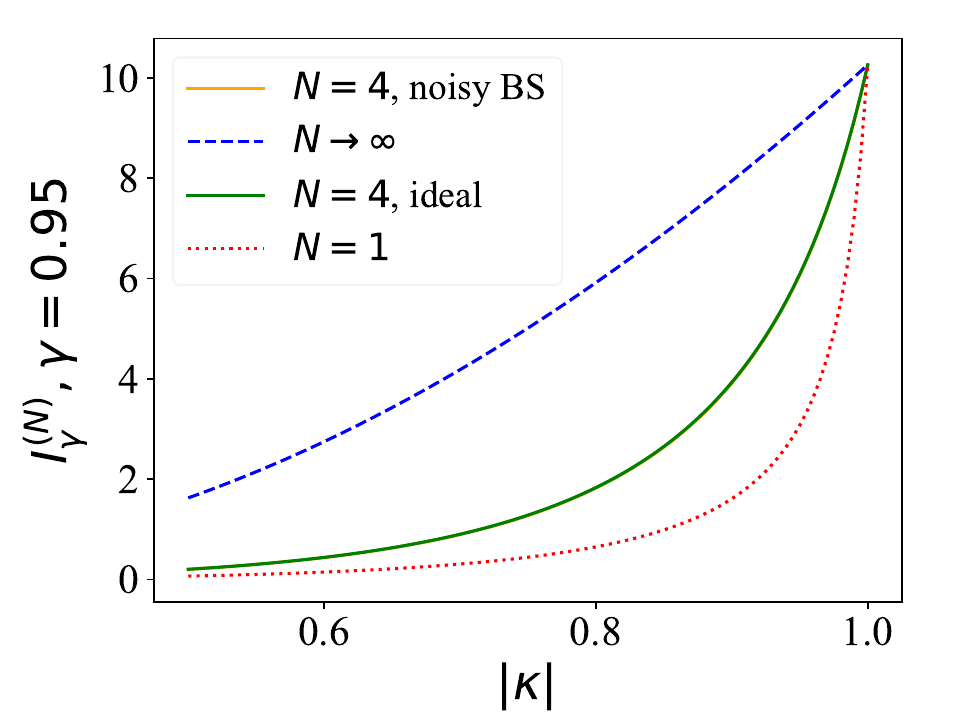}
\caption{\label{fig:uneven_bs} Left: QFI for $\phi$ when $\gamma = 1$ for $N=1,4,\infty$ where the noisy $N=4$ beam splitters are shown in the orange shaded region. Right: QFI for $\gamma$ when $\gamma = 0.95$ for $N=1,4,\infty$ where the reflectivities of the $N=4$ beam splitters are shown in the orange shaded region. Each constituent two-mode beam splitter has reflectivity $r_i $ with added noise with zero mean and $2\%$ standard deviation. Note that the shaded region is barely visible over the size of the plot.} 
\end{figure}

For each reflectivity $r_1, r_2,r_3$, we added zero-mean Gaussian noise with standard deviation of $2\%$ and averaged the QFI for 100 runs. The QFI is shown in Fig.~\ref{fig:uneven_bs} for $|\kappa| \in [0.5,1]$. Only the modes $A_0$ and $B_0$ are considered at the output. It can be seen that the non-ideal reflectivities show little effect on the QFI: the QFI for the noisy beam splitters is shown in orange, where this region is barely visible above the size of the marker.

\subsection{Noisy detectors}
\label{sec:dark_count}

We now analyse the scenario where the detectors being noisy. Note that detector dark counts occurs independently, and does not affect our error mitigation scheme -- our scheme mitigates errors during signal transmission, and detector dark counts occur at the measurement stage.

Let the probability of a dark count be $p$, this can be modelled by depolarization noise,
\begin{align}
\mathcal{E}(\rho) = (1-p)\rho + p \frac{\openone}{d}
\end{align}
where $d$ is the dimension of the system.

After the inverse unitary $\mathcal{F}^{-1} $ (Fig.~\ref{fig:scheme}), the state in modes $A_0$ and $B_0$ is given in Eq.~\eqref{eq:final_rho_prior_detection}. Since these two modes are the only ones containing information on the parameters of interest, these are the ones we need to consider. In this case, we have $d=2$.
Take the (un-normalised) state in Eq.~\eqref{eq:final_rho_prior_detection}
\begin{align}
\varrho &= \left(\begin{array}{cc}
 \frac{1}{N}(1+ (N-1)|\kappa|^2 ) & \gamma e^{i\phi}|\kappa|^2  \\
 \gamma e^{-i\phi}|\kappa|^2 & \frac{1}{N}(1+ (N-1)|\kappa|^2 \\
\end{array}
\right) \nn
\mathcal{E}(\varrho) &= \left(
\begin{array}{cc}
 \frac{\kappa ^2 (N-1)-\left(\kappa ^2-1\right) (N-1) p+1}{2 N} & -\frac{1}{2} \gamma  \kappa ^2 (p-1) e^{i \phi } \\
 -\frac{1}{2} \gamma  \kappa ^2 (p-1) e^{-i \phi } & \frac{\kappa ^2 (N-1)-\left(\kappa ^2-1\right) (N-1) p+1}{2 N} \\
\end{array}
\right)
\end{align}

It follows that the QFI for the two paramter is
\begin{align}
I_\phi^\text{dep} &= \frac{\gamma ^2 \kappa ^4 N (1-p)^2}{1-\left(\kappa ^2 (N-1) (p-1)\right)+N p-p} \nn
I_\gamma^\text{dep} &= \frac{N}{\frac{\gamma ^2 N^2}{\kappa ^2 (N-1) (p-1)-N p+p-1}+\frac{(N-1) p+1}{\kappa ^4 (p-1)^2}+\frac{1-N}{\kappa ^2 (p-1)}} 
\end{align}
and in the limit that $N\rightarrow \infty$
\begin{align}
\lim_{N\rightarrow\infty}I_\phi^\text{dep} &=  \frac{\gamma ^2 \kappa ^4 (1-p)^2}{\kappa ^2+\left(1-\kappa ^2\right) p} \nn
\lim_{N\rightarrow\infty}I_\gamma^\text{dep} &= \frac{\kappa ^4 (p-1)^2 \left(\kappa ^2 (p-1)-p\right)}{-p^2+\left(\gamma ^2-1\right) \kappa ^4 (p-1)^2+2 \kappa ^2 (p-1) p}
\end{align}


\begin{thebibliography}{69}%
\makeatletter
\providecommand \@ifxundefined [1]{%
 \@ifx{#1\undefined}
}%
\providecommand \@ifnum [1]{%
 \ifnum #1\expandafter \@firstoftwo
 \else \expandafter \@secondoftwo
 \fi
}%
\providecommand \@ifx [1]{%
 \ifx #1\expandafter \@firstoftwo
 \else \expandafter \@secondoftwo
 \fi
}%
\providecommand \natexlab [1]{#1}%
\providecommand \enquote  [1]{``#1''}%
\providecommand \bibnamefont  [1]{#1}%
\providecommand \bibfnamefont [1]{#1}%
\providecommand \citenamefont [1]{#1}%
\providecommand \href@noop [0]{\@secondoftwo}%
\providecommand \href [0]{\begingroup \@sanitize@url \@href}%
\providecommand \@href[1]{\@@startlink{#1}\@@href}%
\providecommand \@@href[1]{\endgroup#1\@@endlink}%
\providecommand \@sanitize@url [0]{\catcode `\\12\catcode `\$12\catcode
  `\&12\catcode `\#12\catcode `\^12\catcode `\_12\catcode `\%12\relax}%
\providecommand \@@startlink[1]{}%
\providecommand \@@endlink[0]{}%
\providecommand \url  [0]{\begingroup\@sanitize@url \@url }%
\providecommand \@url [1]{\endgroup\@href {#1}{\urlprefix }}%
\providecommand \urlprefix  [0]{URL }%
\providecommand \Eprint [0]{\href }%
\providecommand \doibase [0]{https://doi.org/}%
\providecommand \selectlanguage [0]{\@gobble}%
\providecommand \bibinfo  [0]{\@secondoftwo}%
\providecommand \bibfield  [0]{\@secondoftwo}%
\providecommand \translation [1]{[#1]}%
\providecommand \BibitemOpen [0]{}%
\providecommand \bibitemStop [0]{}%
\providecommand \bibitemNoStop [0]{.\EOS\space}%
\providecommand \EOS [0]{\spacefactor3000\relax}%
\providecommand \BibitemShut  [1]{\csname bibitem#1\endcsname}%
\let\auto@bib@innerbib\@empty
\bibitem [{\citenamefont {Shor}(1994)}]{Shora}%
  \BibitemOpen
  \bibfield  {author} {\bibinfo {author} {\bibfnamefont {P.}~\bibnamefont
  {Shor}},\ }\bibfield  {title} {\bibinfo {title} {Algorithms for quantum
  computation: discrete logarithms and factoring},\ }in\ \href
  {https://doi.org/10.1109/SFCS.1994.365700} {\emph {\bibinfo {booktitle}
  {Proceedings 35th Annual Symposium on Foundations of Computer Science}}}\
  (\bibinfo {year} {1994})\ pp.\ \bibinfo {pages} {124--134}\BibitemShut
  {NoStop}%
\bibitem [{\citenamefont {Grover}(1996)}]{Grovera}%
  \BibitemOpen
  \bibfield  {author} {\bibinfo {author} {\bibfnamefont {L.~K.}\ \bibnamefont
  {Grover}},\ }\bibfield  {title} {\bibinfo {title} {A fast quantum mechanical
  algorithm for database search},\ }in\ \href
  {https://doi.org/10.1145/237814.237866} {\emph {\bibinfo {booktitle}
  {Proceedings of the Twenty-Eighth Annual ACM Symposium on Theory of
  Computing}}},\ \bibinfo {series and number} {STOC '96}\ (\bibinfo
  {publisher} {Association for Computing Machinery},\ \bibinfo {address} {New
  York, NY, USA},\ \bibinfo {year} {1996})\ p.\ \bibinfo {pages}
  {212–219}\BibitemShut {NoStop}%
\bibitem [{\citenamefont {McArdle}\ \emph {et~al.}(2020)\citenamefont
  {McArdle}, \citenamefont {Endo}, \citenamefont {Aspuru-Guzik}, \citenamefont
  {Benjamin},\ and\ \citenamefont {Yuan}}]{QChem}%
  \BibitemOpen
  \bibfield  {author} {\bibinfo {author} {\bibfnamefont {S.}~\bibnamefont
  {McArdle}}, \bibinfo {author} {\bibfnamefont {S.}~\bibnamefont {Endo}},
  \bibinfo {author} {\bibfnamefont {A.}~\bibnamefont {Aspuru-Guzik}}, \bibinfo
  {author} {\bibfnamefont {S.~C.}\ \bibnamefont {Benjamin}},\ and\ \bibinfo
  {author} {\bibfnamefont {X.}~\bibnamefont {Yuan}},\ }\bibfield  {title}
  {\bibinfo {title} {Quantum computational chemistry},\ }\href
  {https://doi.org/10.1103/RevModPhys.92.015003} {\bibfield  {journal}
  {\bibinfo  {journal} {Rev. Mod. Phys.}\ }\textbf {\bibinfo {volume} {92}},\
  \bibinfo {pages} {015003} (\bibinfo {year} {2020})}\BibitemShut {NoStop}%
\bibitem [{\citenamefont {Bañuls}\ \emph {et~al.}(2020)\citenamefont
  {Bañuls}, \citenamefont {Blatt}, \citenamefont {Catani}, \citenamefont
  {Celi}, \citenamefont {Cirac}, \citenamefont {Dalmonte}, \citenamefont
  {Fallani}, \citenamefont {Jansen}, \citenamefont {Lewenstein}, \citenamefont
  {Montangero}, \citenamefont {Muschik}, \citenamefont {Reznik}, \citenamefont
  {Rico}, \citenamefont {Tagliacozzo}, \citenamefont {Acoleyen}, \citenamefont
  {Verstraete}, \citenamefont {Wiese}, \citenamefont {Wingate}, \citenamefont
  {Zakrzewski},\ and\ \citenamefont {Zoller}}]{gauge}%
  \BibitemOpen
  \bibfield  {author} {\bibinfo {author} {\bibfnamefont {M.~C.}\ \bibnamefont
  {Bañuls}}, \bibinfo {author} {\bibfnamefont {R.}~\bibnamefont {Blatt}},
  \bibinfo {author} {\bibfnamefont {J.}~\bibnamefont {Catani}}, \bibinfo
  {author} {\bibfnamefont {A.}~\bibnamefont {Celi}}, \bibinfo {author}
  {\bibfnamefont {J.~I.}\ \bibnamefont {Cirac}}, \bibinfo {author}
  {\bibfnamefont {M.}~\bibnamefont {Dalmonte}}, \bibinfo {author}
  {\bibfnamefont {L.}~\bibnamefont {Fallani}}, \bibinfo {author} {\bibfnamefont
  {K.}~\bibnamefont {Jansen}}, \bibinfo {author} {\bibfnamefont
  {M.}~\bibnamefont {Lewenstein}}, \bibinfo {author} {\bibfnamefont
  {S.}~\bibnamefont {Montangero}}, \bibinfo {author} {\bibfnamefont {C.~A.}\
  \bibnamefont {Muschik}}, \bibinfo {author} {\bibfnamefont {B.}~\bibnamefont
  {Reznik}}, \bibinfo {author} {\bibfnamefont {E.}~\bibnamefont {Rico}},
  \bibinfo {author} {\bibfnamefont {L.}~\bibnamefont {Tagliacozzo}}, \bibinfo
  {author} {\bibfnamefont {K.~V.}\ \bibnamefont {Acoleyen}}, \bibinfo {author}
  {\bibfnamefont {F.}~\bibnamefont {Verstraete}}, \bibinfo {author}
  {\bibfnamefont {U.-J.}\ \bibnamefont {Wiese}}, \bibinfo {author}
  {\bibfnamefont {M.}~\bibnamefont {Wingate}}, \bibinfo {author} {\bibfnamefont
  {J.}~\bibnamefont {Zakrzewski}},\ and\ \bibinfo {author} {\bibfnamefont
  {P.}~\bibnamefont {Zoller}},\ }\bibfield  {title} {\bibinfo {title}
  {Simulating lattice gauge theories within quantum technologies},\ }\href
  {https://doi.org/10.1140/epjd/e2020-100571-8} {\bibfield  {journal} {\bibinfo
   {journal} {Eur. Phys. J. D}\ }\textbf {\bibinfo {volume} {74}},\ \bibinfo
  {pages} {165} (\bibinfo {year} {2020})}\BibitemShut {NoStop}%
\bibitem [{\citenamefont {Di~Meglio}\ \emph {et~al.}(2023)\citenamefont
  {Di~Meglio}, \citenamefont {Jansen}, \citenamefont {Tavernelli},
  \citenamefont {Alexandrou}, \citenamefont {Arunachalam}, \citenamefont
  {Bauer}, \citenamefont {Borras}, \citenamefont {Carrazza}, \citenamefont
  {Crippa}, \citenamefont {Croft}, \citenamefont {de~Putter}, \citenamefont
  {Delgado}, \citenamefont {Dunjko}, \citenamefont {Egger}, \citenamefont
  {Fernandez-Combarro}, \citenamefont {Fuchs}, \citenamefont {Funcke},
  \citenamefont {Gonzalez-Cuadra}, \citenamefont {Grossi}, \citenamefont
  {Halimeh}, \citenamefont {Holmes}, \citenamefont {Kuhn}, \citenamefont
  {Lacroix}, \citenamefont {Lewis}, \citenamefont {Lucchesi}, \citenamefont
  {Martinez}, \citenamefont {Meloni}, \citenamefont {Mezzacapo}, \citenamefont
  {Montangero}, \citenamefont {Nagano}, \citenamefont {Radescu}, \citenamefont
  {Ortega}, \citenamefont {Roggero}, \citenamefont {Schuhmacher}, \citenamefont
  {Seixas}, \citenamefont {Silvi}, \citenamefont {Spentzouris}, \citenamefont
  {Tacchino}, \citenamefont {Temme}, \citenamefont {Terashi}, \citenamefont
  {Tura}, \citenamefont {Tuysuz}, \citenamefont {Vallecorsa}, \citenamefont
  {Wiese}, \citenamefont {Yoo},\ and\ \citenamefont {Zhang}}]{QHEP}%
  \BibitemOpen
  \bibfield  {author} {\bibinfo {author} {\bibfnamefont {A.}~\bibnamefont
  {Di~Meglio}}, \bibinfo {author} {\bibfnamefont {K.}~\bibnamefont {Jansen}},
  \bibinfo {author} {\bibfnamefont {I.}~\bibnamefont {Tavernelli}}, \bibinfo
  {author} {\bibfnamefont {C.}~\bibnamefont {Alexandrou}}, \bibinfo {author}
  {\bibfnamefont {S.}~\bibnamefont {Arunachalam}}, \bibinfo {author}
  {\bibfnamefont {C.~W.}\ \bibnamefont {Bauer}}, \bibinfo {author}
  {\bibfnamefont {K.}~\bibnamefont {Borras}}, \bibinfo {author} {\bibfnamefont
  {S.}~\bibnamefont {Carrazza}}, \bibinfo {author} {\bibfnamefont
  {A.}~\bibnamefont {Crippa}}, \bibinfo {author} {\bibfnamefont
  {V.}~\bibnamefont {Croft}}, \bibinfo {author} {\bibfnamefont
  {R.}~\bibnamefont {de~Putter}}, \bibinfo {author} {\bibfnamefont
  {A.}~\bibnamefont {Delgado}}, \bibinfo {author} {\bibfnamefont
  {V.}~\bibnamefont {Dunjko}}, \bibinfo {author} {\bibfnamefont {D.~J.}\
  \bibnamefont {Egger}}, \bibinfo {author} {\bibfnamefont {E.}~\bibnamefont
  {Fernandez-Combarro}}, \bibinfo {author} {\bibfnamefont {E.}~\bibnamefont
  {Fuchs}}, \bibinfo {author} {\bibfnamefont {L.}~\bibnamefont {Funcke}},
  \bibinfo {author} {\bibfnamefont {D.}~\bibnamefont {Gonzalez-Cuadra}},
  \bibinfo {author} {\bibfnamefont {M.}~\bibnamefont {Grossi}}, \bibinfo
  {author} {\bibfnamefont {J.~C.}\ \bibnamefont {Halimeh}}, \bibinfo {author}
  {\bibfnamefont {Z.}~\bibnamefont {Holmes}}, \bibinfo {author} {\bibfnamefont
  {S.}~\bibnamefont {Kuhn}}, \bibinfo {author} {\bibfnamefont {D.}~\bibnamefont
  {Lacroix}}, \bibinfo {author} {\bibfnamefont {R.}~\bibnamefont {Lewis}},
  \bibinfo {author} {\bibfnamefont {D.}~\bibnamefont {Lucchesi}}, \bibinfo
  {author} {\bibfnamefont {M.~L.}\ \bibnamefont {Martinez}}, \bibinfo {author}
  {\bibfnamefont {F.}~\bibnamefont {Meloni}}, \bibinfo {author} {\bibfnamefont
  {A.}~\bibnamefont {Mezzacapo}}, \bibinfo {author} {\bibfnamefont
  {S.}~\bibnamefont {Montangero}}, \bibinfo {author} {\bibfnamefont
  {L.}~\bibnamefont {Nagano}}, \bibinfo {author} {\bibfnamefont
  {V.}~\bibnamefont {Radescu}}, \bibinfo {author} {\bibfnamefont {E.~R.}\
  \bibnamefont {Ortega}}, \bibinfo {author} {\bibfnamefont {A.}~\bibnamefont
  {Roggero}}, \bibinfo {author} {\bibfnamefont {J.}~\bibnamefont
  {Schuhmacher}}, \bibinfo {author} {\bibfnamefont {J.}~\bibnamefont {Seixas}},
  \bibinfo {author} {\bibfnamefont {P.}~\bibnamefont {Silvi}}, \bibinfo
  {author} {\bibfnamefont {P.}~\bibnamefont {Spentzouris}}, \bibinfo {author}
  {\bibfnamefont {F.}~\bibnamefont {Tacchino}}, \bibinfo {author}
  {\bibfnamefont {K.}~\bibnamefont {Temme}}, \bibinfo {author} {\bibfnamefont
  {K.}~\bibnamefont {Terashi}}, \bibinfo {author} {\bibfnamefont
  {J.}~\bibnamefont {Tura}}, \bibinfo {author} {\bibfnamefont {C.}~\bibnamefont
  {Tuysuz}}, \bibinfo {author} {\bibfnamefont {S.}~\bibnamefont {Vallecorsa}},
  \bibinfo {author} {\bibfnamefont {U.-J.}\ \bibnamefont {Wiese}}, \bibinfo
  {author} {\bibfnamefont {S.}~\bibnamefont {Yoo}},\ and\ \bibinfo {author}
  {\bibfnamefont {J.}~\bibnamefont {Zhang}},\ }\bibfield  {title} {\bibinfo
  {title} {Quantum computing for high-energy physics: State of the art and
  challenges. summary of the qc4hep working group},\ }\href@noop {} {\bibfield
  {journal} {\bibinfo  {journal} {arXiv preprint arXiv:2307.03236}\ } (\bibinfo
  {year} {2023})}\BibitemShut {NoStop}%
\bibitem [{\citenamefont {Boto}\ \emph {et~al.}(2000)\citenamefont {Boto},
  \citenamefont {Kok}, \citenamefont {Abrams}, \citenamefont {Braunstein},
  \citenamefont {Williams},\ and\ \citenamefont {Dowling}}]{Dowling}%
  \BibitemOpen
  \bibfield  {author} {\bibinfo {author} {\bibfnamefont {A.~N.}\ \bibnamefont
  {Boto}}, \bibinfo {author} {\bibfnamefont {P.}~\bibnamefont {Kok}}, \bibinfo
  {author} {\bibfnamefont {D.~S.}\ \bibnamefont {Abrams}}, \bibinfo {author}
  {\bibfnamefont {S.~L.}\ \bibnamefont {Braunstein}}, \bibinfo {author}
  {\bibfnamefont {C.~P.}\ \bibnamefont {Williams}},\ and\ \bibinfo {author}
  {\bibfnamefont {J.~P.}\ \bibnamefont {Dowling}},\ }\bibfield  {title}
  {\bibinfo {title} {Quantum interferometric optical lithography: Exploiting
  entanglement to beat the diffraction limit},\ }\href
  {https://doi.org/10.1103/PhysRevLett.85.2733} {\bibfield  {journal} {\bibinfo
   {journal} {Phys. Rev. Lett.}\ }\textbf {\bibinfo {volume} {85}},\ \bibinfo
  {pages} {2733} (\bibinfo {year} {2000})}\BibitemShut {NoStop}%
\bibitem [{\citenamefont {Dowling}(2008)}]{JPD}%
  \BibitemOpen
  \bibfield  {author} {\bibinfo {author} {\bibfnamefont {J.~P.}\ \bibnamefont
  {Dowling}},\ }\bibfield  {title} {\bibinfo {title} {Quantum optical
  metrology-- the lowdown on high-n00n states},\ }\href
  {https://doi.org/10.1080/00107510802091298} {\bibfield  {journal} {\bibinfo
  {journal} {Contemporary Physics}\ }\textbf {\bibinfo {volume} {49}},\
  \bibinfo {pages} {125} (\bibinfo {year} {2008})},\ \Eprint
  {https://arxiv.org/abs/https://doi.org/10.1080/00107510802091298}
  {https://doi.org/10.1080/00107510802091298} \BibitemShut {NoStop}%
\bibitem [{\citenamefont {Gottesman}\ \emph {et~al.}(2012)\citenamefont
  {Gottesman}, \citenamefont {Jennewein},\ and\ \citenamefont
  {Croke}}]{PhysRevLett.109.070503}%
  \BibitemOpen
  \bibfield  {author} {\bibinfo {author} {\bibfnamefont {D.}~\bibnamefont
  {Gottesman}}, \bibinfo {author} {\bibfnamefont {T.}~\bibnamefont
  {Jennewein}},\ and\ \bibinfo {author} {\bibfnamefont {S.}~\bibnamefont
  {Croke}},\ }\bibfield  {title} {\bibinfo {title} {Longer-baseline telescopes
  using quantum repeaters},\ }\href
  {https://doi.org/10.1103/PhysRevLett.109.070503} {\bibfield  {journal}
  {\bibinfo  {journal} {Phys. Rev. Lett.}\ }\textbf {\bibinfo {volume} {109}},\
  \bibinfo {pages} {070503} (\bibinfo {year} {2012})}\BibitemShut {NoStop}%
\bibitem [{\citenamefont {Tsang}\ \emph {et~al.}(2016)\citenamefont {Tsang},
  \citenamefont {Nair},\ and\ \citenamefont {Lu}}]{Tsang}%
  \BibitemOpen
  \bibfield  {author} {\bibinfo {author} {\bibfnamefont {M.}~\bibnamefont
  {Tsang}}, \bibinfo {author} {\bibfnamefont {R.}~\bibnamefont {Nair}},\ and\
  \bibinfo {author} {\bibfnamefont {X.-M.}\ \bibnamefont {Lu}},\ }\bibfield
  {title} {\bibinfo {title} {Quantum theory of superresolution for two
  incoherent optical point sources},\ }\href
  {https://doi.org/10.1103/PhysRevX.6.031033} {\bibfield  {journal} {\bibinfo
  {journal} {Phys. Rev. X}\ }\textbf {\bibinfo {volume} {6}},\ \bibinfo {pages}
  {031033} (\bibinfo {year} {2016})}\BibitemShut {NoStop}%
\bibitem [{\citenamefont {Degen}\ \emph {et~al.}(2017)\citenamefont {Degen},
  \citenamefont {Reinhard},\ and\ \citenamefont {Cappellaro}}]{Cappellaro}%
  \BibitemOpen
  \bibfield  {author} {\bibinfo {author} {\bibfnamefont {C.~L.}\ \bibnamefont
  {Degen}}, \bibinfo {author} {\bibfnamefont {F.}~\bibnamefont {Reinhard}},\
  and\ \bibinfo {author} {\bibfnamefont {P.}~\bibnamefont {Cappellaro}},\
  }\bibfield  {title} {\bibinfo {title} {Quantum sensing},\ }\href
  {https://doi.org/10.1103/RevModPhys.89.035002} {\bibfield  {journal}
  {\bibinfo  {journal} {Rev. Mod. Phys.}\ }\textbf {\bibinfo {volume} {89}},\
  \bibinfo {pages} {035002} (\bibinfo {year} {2017})}\BibitemShut {NoStop}%
\bibitem [{\citenamefont {Kok}\ \emph {et~al.}(2001)\citenamefont {Kok},
  \citenamefont {Boto}, \citenamefont {Abrams}, \citenamefont {Williams},
  \citenamefont {Braunstein},\ and\ \citenamefont {Dowling}}]{Kok}%
  \BibitemOpen
  \bibfield  {author} {\bibinfo {author} {\bibfnamefont {P.}~\bibnamefont
  {Kok}}, \bibinfo {author} {\bibfnamefont {A.~N.}\ \bibnamefont {Boto}},
  \bibinfo {author} {\bibfnamefont {D.~S.}\ \bibnamefont {Abrams}}, \bibinfo
  {author} {\bibfnamefont {C.~P.}\ \bibnamefont {Williams}}, \bibinfo {author}
  {\bibfnamefont {S.~L.}\ \bibnamefont {Braunstein}},\ and\ \bibinfo {author}
  {\bibfnamefont {J.~P.}\ \bibnamefont {Dowling}},\ }\bibfield  {title}
  {\bibinfo {title} {Quantum-interferometric optical lithography: Towards
  arbitrary two-dimensional patterns},\ }\href
  {https://doi.org/10.1103/PhysRevA.63.063407} {\bibfield  {journal} {\bibinfo
  {journal} {Phys. Rev. A}\ }\textbf {\bibinfo {volume} {63}},\ \bibinfo
  {pages} {063407} (\bibinfo {year} {2001})}\BibitemShut {NoStop}%
\bibitem [{\citenamefont {Giovannetti}\ \emph {et~al.}(2009)\citenamefont
  {Giovannetti}, \citenamefont {Lloyd}, \citenamefont {Maccone},\ and\
  \citenamefont {Shapiro}}]{Giovannetti}%
  \BibitemOpen
  \bibfield  {author} {\bibinfo {author} {\bibfnamefont {V.}~\bibnamefont
  {Giovannetti}}, \bibinfo {author} {\bibfnamefont {S.}~\bibnamefont {Lloyd}},
  \bibinfo {author} {\bibfnamefont {L.}~\bibnamefont {Maccone}},\ and\ \bibinfo
  {author} {\bibfnamefont {J.~H.}\ \bibnamefont {Shapiro}},\ }\bibfield
  {title} {\bibinfo {title} {Sub-rayleigh-diffraction-bound quantum imaging},\
  }\href {https://doi.org/10.1103/PhysRevA.79.013827} {\bibfield  {journal}
  {\bibinfo  {journal} {Phys. Rev. A}\ }\textbf {\bibinfo {volume} {79}},\
  \bibinfo {pages} {013827} (\bibinfo {year} {2009})}\BibitemShut {NoStop}%
\bibitem [{\citenamefont {Abbattista}\ \emph {et~al.}(2021)\citenamefont
  {Abbattista}, \citenamefont {Amoruso}, \citenamefont {Burri}, \citenamefont
  {Charbon}, \citenamefont {Di~Lena}, \citenamefont {Garuccio}, \citenamefont
  {Giannella}, \citenamefont {Hradil}, \citenamefont {Iacobellis},
  \citenamefont {Massaro}, \citenamefont {Mos}, \citenamefont {Motka},
  \citenamefont {Paúr}, \citenamefont {Pepe}, \citenamefont {Peterek},
  \citenamefont {Petrelli}, \citenamefont {Řeháček}, \citenamefont
  {Santoro}, \citenamefont {Scattarella}, \citenamefont {Ulku}, \citenamefont
  {Vasiukov}, \citenamefont {Wayne}, \citenamefont {Bruschini}, \citenamefont
  {D’Angelo}, \citenamefont {Ieronymaki},\ and\ \citenamefont
  {Stoklasa}}]{app11146414}%
  \BibitemOpen
  \bibfield  {author} {\bibinfo {author} {\bibfnamefont {C.}~\bibnamefont
  {Abbattista}}, \bibinfo {author} {\bibfnamefont {L.}~\bibnamefont {Amoruso}},
  \bibinfo {author} {\bibfnamefont {S.}~\bibnamefont {Burri}}, \bibinfo
  {author} {\bibfnamefont {E.}~\bibnamefont {Charbon}}, \bibinfo {author}
  {\bibfnamefont {F.}~\bibnamefont {Di~Lena}}, \bibinfo {author} {\bibfnamefont
  {A.}~\bibnamefont {Garuccio}}, \bibinfo {author} {\bibfnamefont
  {D.}~\bibnamefont {Giannella}}, \bibinfo {author} {\bibfnamefont
  {Z.}~\bibnamefont {Hradil}}, \bibinfo {author} {\bibfnamefont
  {M.}~\bibnamefont {Iacobellis}}, \bibinfo {author} {\bibfnamefont
  {G.}~\bibnamefont {Massaro}}, \bibinfo {author} {\bibfnamefont
  {P.}~\bibnamefont {Mos}}, \bibinfo {author} {\bibfnamefont {L.}~\bibnamefont
  {Motka}}, \bibinfo {author} {\bibfnamefont {M.}~\bibnamefont {Paúr}},
  \bibinfo {author} {\bibfnamefont {F.~V.}\ \bibnamefont {Pepe}}, \bibinfo
  {author} {\bibfnamefont {M.}~\bibnamefont {Peterek}}, \bibinfo {author}
  {\bibfnamefont {I.}~\bibnamefont {Petrelli}}, \bibinfo {author}
  {\bibfnamefont {J.}~\bibnamefont {Řeháček}}, \bibinfo {author}
  {\bibfnamefont {F.}~\bibnamefont {Santoro}}, \bibinfo {author} {\bibfnamefont
  {F.}~\bibnamefont {Scattarella}}, \bibinfo {author} {\bibfnamefont
  {A.}~\bibnamefont {Ulku}}, \bibinfo {author} {\bibfnamefont {S.}~\bibnamefont
  {Vasiukov}}, \bibinfo {author} {\bibfnamefont {M.}~\bibnamefont {Wayne}},
  \bibinfo {author} {\bibfnamefont {C.}~\bibnamefont {Bruschini}}, \bibinfo
  {author} {\bibfnamefont {M.}~\bibnamefont {D’Angelo}}, \bibinfo {author}
  {\bibfnamefont {M.}~\bibnamefont {Ieronymaki}},\ and\ \bibinfo {author}
  {\bibfnamefont {B.}~\bibnamefont {Stoklasa}},\ }\bibfield  {title} {\bibinfo
  {title} {Towards quantum 3d imaging devices},\ }\bibfield  {journal}
  {\bibinfo  {journal} {Applied Sciences}\ }\textbf {\bibinfo {volume} {11}},\
  \href {https://doi.org/10.3390/app11146414} {10.3390/app11146414} (\bibinfo
  {year} {2021})\BibitemShut {NoStop}%
\bibitem [{\citenamefont {Bennett}\ and\ \citenamefont
  {Brassard}(1984)}]{BB84}%
  \BibitemOpen
  \bibfield  {author} {\bibinfo {author} {\bibfnamefont {C.~H.}\ \bibnamefont
  {Bennett}}\ and\ \bibinfo {author} {\bibfnamefont {G.}~\bibnamefont
  {Brassard}},\ }\bibfield  {title} {\bibinfo {title} {Quantum cryptography:
  Public key distribution and coin tossing},\ }\href@noop {} {\bibfield
  {journal} {\bibinfo  {journal} {Proceedings of the IEEE International
  Conference on Computers, Systems and Signal Processing, Bangalore, India,
  10–12 December, 1984}\ }\textbf {\bibinfo {volume} {175}},\ \bibinfo
  {pages} {8} (\bibinfo {year} {1984})}\BibitemShut {NoStop}%
\bibitem [{\citenamefont {Ekert}(1991)}]{Ekert}%
  \BibitemOpen
  \bibfield  {author} {\bibinfo {author} {\bibfnamefont {A.~K.}\ \bibnamefont
  {Ekert}},\ }\bibfield  {title} {\bibinfo {title} {Quantum cryptography based
  on bell's theorem},\ }\href {https://doi.org/10.1103/PhysRevLett.67.661}
  {\bibfield  {journal} {\bibinfo  {journal} {Phys. Rev. Lett.}\ }\textbf
  {\bibinfo {volume} {67}},\ \bibinfo {pages} {661} (\bibinfo {year}
  {1991})}\BibitemShut {NoStop}%
\bibitem [{\citenamefont {Ac\'{\i}n}\ \emph {et~al.}(2007)\citenamefont
  {Ac\'{\i}n}, \citenamefont {Brunner}, \citenamefont {Gisin}, \citenamefont
  {Massar}, \citenamefont {Pironio},\ and\ \citenamefont
  {Scarani}}]{AcinPRL2007}%
  \BibitemOpen
  \bibfield  {author} {\bibinfo {author} {\bibfnamefont {A.}~\bibnamefont
  {Ac\'{\i}n}}, \bibinfo {author} {\bibfnamefont {N.}~\bibnamefont {Brunner}},
  \bibinfo {author} {\bibfnamefont {N.}~\bibnamefont {Gisin}}, \bibinfo
  {author} {\bibfnamefont {S.}~\bibnamefont {Massar}}, \bibinfo {author}
  {\bibfnamefont {S.}~\bibnamefont {Pironio}},\ and\ \bibinfo {author}
  {\bibfnamefont {V.}~\bibnamefont {Scarani}},\ }\bibfield  {title} {\bibinfo
  {title} {Device-independent security of quantum cryptography against
  collective attacks},\ }\href {https://doi.org/10.1103/PhysRevLett.98.230501}
  {\bibfield  {journal} {\bibinfo  {journal} {Phys. Rev. Lett.}\ }\textbf
  {\bibinfo {volume} {98}},\ \bibinfo {pages} {230501} (\bibinfo {year}
  {2007})}\BibitemShut {NoStop}%
\bibitem [{\citenamefont {Bremner}\ \emph {et~al.}(2017)\citenamefont
  {Bremner}, \citenamefont {Montanaro},\ and\ \citenamefont
  {Shepherd}}]{bremner2017achieving}%
  \BibitemOpen
  \bibfield  {author} {\bibinfo {author} {\bibfnamefont {M.~J.}\ \bibnamefont
  {Bremner}}, \bibinfo {author} {\bibfnamefont {A.}~\bibnamefont {Montanaro}},\
  and\ \bibinfo {author} {\bibfnamefont {D.~J.}\ \bibnamefont {Shepherd}},\
  }\bibfield  {title} {\bibinfo {title} {Achieving quantum supremacy with
  sparse and noisy commuting quantum computations},\ }\href@noop {} {\bibfield
  {journal} {\bibinfo  {journal} {Quantum}\ }\textbf {\bibinfo {volume} {1}},\
  \bibinfo {pages} {8} (\bibinfo {year} {2017})}\BibitemShut {NoStop}%
\bibitem [{\citenamefont {Noh}\ \emph {et~al.}(2020)\citenamefont {Noh},
  \citenamefont {Jiang},\ and\ \citenamefont
  {Fefferman}}]{Noh2020efficientclassical}%
  \BibitemOpen
  \bibfield  {author} {\bibinfo {author} {\bibfnamefont {K.}~\bibnamefont
  {Noh}}, \bibinfo {author} {\bibfnamefont {L.}~\bibnamefont {Jiang}},\ and\
  \bibinfo {author} {\bibfnamefont {B.}~\bibnamefont {Fefferman}},\ }\bibfield
  {title} {\bibinfo {title} {Efficient classical simulation of noisy random
  quantum circuits in one dimension},\ }\href
  {https://doi.org/10.22331/q-2020-09-11-318} {\bibfield  {journal} {\bibinfo
  {journal} {{Quantum}}\ }\textbf {\bibinfo {volume} {4}},\ \bibinfo {pages}
  {318} (\bibinfo {year} {2020})}\BibitemShut {NoStop}%
\bibitem [{\citenamefont {Aharonov}\ \emph {et~al.}(2023)\citenamefont
  {Aharonov}, \citenamefont {Gao}, \citenamefont {Landau}, \citenamefont
  {Liu},\ and\ \citenamefont {Vazirani}}]{aharonov2023polynomial}%
  \BibitemOpen
  \bibfield  {author} {\bibinfo {author} {\bibfnamefont {D.}~\bibnamefont
  {Aharonov}}, \bibinfo {author} {\bibfnamefont {X.}~\bibnamefont {Gao}},
  \bibinfo {author} {\bibfnamefont {Z.}~\bibnamefont {Landau}}, \bibinfo
  {author} {\bibfnamefont {Y.}~\bibnamefont {Liu}},\ and\ \bibinfo {author}
  {\bibfnamefont {U.}~\bibnamefont {Vazirani}},\ }\bibfield  {title} {\bibinfo
  {title} {A polynomial-time classical algorithm for noisy random circuit
  sampling},\ }in\ \href@noop {} {\emph {\bibinfo {booktitle} {Proceedings of
  the 55th Annual ACM Symposium on Theory of Computing}}}\ (\bibinfo {year}
  {2023})\ pp.\ \bibinfo {pages} {945--957}\BibitemShut {NoStop}%
\bibitem [{\citenamefont {Terhal}(2015)}]{RevModPhys.87.307}%
  \BibitemOpen
  \bibfield  {author} {\bibinfo {author} {\bibfnamefont {B.~M.}\ \bibnamefont
  {Terhal}},\ }\bibfield  {title} {\bibinfo {title} {Quantum error correction
  for quantum memories},\ }\href {https://doi.org/10.1103/RevModPhys.87.307}
  {\bibfield  {journal} {\bibinfo  {journal} {Rev. Mod. Phys.}\ }\textbf
  {\bibinfo {volume} {87}},\ \bibinfo {pages} {307} (\bibinfo {year}
  {2015})}\BibitemShut {NoStop}%
\bibitem [{\citenamefont {Roffe}(2019)}]{roffe2019quantum}%
  \BibitemOpen
  \bibfield  {author} {\bibinfo {author} {\bibfnamefont {J.}~\bibnamefont
  {Roffe}},\ }\bibfield  {title} {\bibinfo {title} {Quantum error correction:
  an introductory guide},\ }\href@noop {} {\bibfield  {journal} {\bibinfo
  {journal} {Contemporary Physics}\ }\textbf {\bibinfo {volume} {60}},\
  \bibinfo {pages} {226} (\bibinfo {year} {2019})}\BibitemShut {NoStop}%
\bibitem [{\citenamefont {Devitt}\ \emph {et~al.}(2013)\citenamefont {Devitt},
  \citenamefont {Munro},\ and\ \citenamefont {Nemoto}}]{devitt2013quantum}%
  \BibitemOpen
  \bibfield  {author} {\bibinfo {author} {\bibfnamefont {S.~J.}\ \bibnamefont
  {Devitt}}, \bibinfo {author} {\bibfnamefont {W.~J.}\ \bibnamefont {Munro}},\
  and\ \bibinfo {author} {\bibfnamefont {K.}~\bibnamefont {Nemoto}},\
  }\bibfield  {title} {\bibinfo {title} {Quantum error correction for
  beginners},\ }\href@noop {} {\bibfield  {journal} {\bibinfo  {journal}
  {Reports on Progress in Physics}\ }\textbf {\bibinfo {volume} {76}},\
  \bibinfo {pages} {076001} (\bibinfo {year} {2013})}\BibitemShut {NoStop}%
\bibitem [{\citenamefont {Campbell}\ \emph {et~al.}(2017)\citenamefont
  {Campbell}, \citenamefont {Terhal},\ and\ \citenamefont
  {Vuillot}}]{campbell2017roads}%
  \BibitemOpen
  \bibfield  {author} {\bibinfo {author} {\bibfnamefont {E.~T.}\ \bibnamefont
  {Campbell}}, \bibinfo {author} {\bibfnamefont {B.~M.}\ \bibnamefont
  {Terhal}},\ and\ \bibinfo {author} {\bibfnamefont {C.}~\bibnamefont
  {Vuillot}},\ }\bibfield  {title} {\bibinfo {title} {Roads towards
  fault-tolerant universal quantum computation},\ }\href@noop {} {\bibfield
  {journal} {\bibinfo  {journal} {Nature}\ }\textbf {\bibinfo {volume} {549}},\
  \bibinfo {pages} {172} (\bibinfo {year} {2017})}\BibitemShut {NoStop}%
\bibitem [{\citenamefont {Preskill}(2018)}]{preskill2018quantum}%
  \BibitemOpen
  \bibfield  {author} {\bibinfo {author} {\bibfnamefont {J.}~\bibnamefont
  {Preskill}},\ }\bibfield  {title} {\bibinfo {title} {Quantum computing in the
  nisq era and beyond},\ }\href@noop {} {\bibfield  {journal} {\bibinfo
  {journal} {Quantum}\ }\textbf {\bibinfo {volume} {2}},\ \bibinfo {pages} {79}
  (\bibinfo {year} {2018})}\BibitemShut {NoStop}%
\bibitem [{\citenamefont {Bultrini}\ \emph {et~al.}(2023)\citenamefont
  {Bultrini}, \citenamefont {Gordon}, \citenamefont {Czarnik}, \citenamefont
  {Arrasmith}, \citenamefont {Cerezo}, \citenamefont {Coles},\ and\
  \citenamefont {Cincio}}]{Bultrini2023unifying}%
  \BibitemOpen
  \bibfield  {author} {\bibinfo {author} {\bibfnamefont {D.}~\bibnamefont
  {Bultrini}}, \bibinfo {author} {\bibfnamefont {M.~H.}\ \bibnamefont
  {Gordon}}, \bibinfo {author} {\bibfnamefont {P.}~\bibnamefont {Czarnik}},
  \bibinfo {author} {\bibfnamefont {A.}~\bibnamefont {Arrasmith}}, \bibinfo
  {author} {\bibfnamefont {M.}~\bibnamefont {Cerezo}}, \bibinfo {author}
  {\bibfnamefont {P.~J.}\ \bibnamefont {Coles}},\ and\ \bibinfo {author}
  {\bibfnamefont {L.}~\bibnamefont {Cincio}},\ }\bibfield  {title} {\bibinfo
  {title} {Unifying and benchmarking state-of-the-art quantum error mitigation
  techniques},\ }\href {https://doi.org/10.22331/q-2023-06-06-1034} {\bibfield
  {journal} {\bibinfo  {journal} {{Quantum}}\ }\textbf {\bibinfo {volume}
  {7}},\ \bibinfo {pages} {1034} (\bibinfo {year} {2023})}\BibitemShut
  {NoStop}%
\bibitem [{\citenamefont {Cai}\ \emph {et~al.}(2022)\citenamefont {Cai},
  \citenamefont {Babbush}, \citenamefont {Benjamin}, \citenamefont {Endo},
  \citenamefont {Huggins}, \citenamefont {Li}, \citenamefont {McClean},\ and\
  \citenamefont {O'Brien}}]{qemit}%
  \BibitemOpen
  \bibfield  {author} {\bibinfo {author} {\bibfnamefont {Z.}~\bibnamefont
  {Cai}}, \bibinfo {author} {\bibfnamefont {R.}~\bibnamefont {Babbush}},
  \bibinfo {author} {\bibfnamefont {S.~C.}\ \bibnamefont {Benjamin}}, \bibinfo
  {author} {\bibfnamefont {S.}~\bibnamefont {Endo}}, \bibinfo {author}
  {\bibfnamefont {W.~J.}\ \bibnamefont {Huggins}}, \bibinfo {author}
  {\bibfnamefont {Y.}~\bibnamefont {Li}}, \bibinfo {author} {\bibfnamefont
  {J.~R.}\ \bibnamefont {McClean}},\ and\ \bibinfo {author} {\bibfnamefont
  {T.~E.}\ \bibnamefont {O'Brien}},\ }\bibfield  {title} {\bibinfo {title}
  {Quantum error mitigation},\ }\href@noop {} {\bibfield  {journal} {\bibinfo
  {journal} {arXiv preprint arXiv:2210.00921}\ } (\bibinfo {year}
  {2022})}\BibitemShut {NoStop}%
\bibitem [{\citenamefont {Gisin}\ \emph {et~al.}(2005)\citenamefont {Gisin},
  \citenamefont {Linden}, \citenamefont {Massar},\ and\ \citenamefont
  {Popescu}}]{PhysRevA.72.012338}%
  \BibitemOpen
  \bibfield  {author} {\bibinfo {author} {\bibfnamefont {N.}~\bibnamefont
  {Gisin}}, \bibinfo {author} {\bibfnamefont {N.}~\bibnamefont {Linden}},
  \bibinfo {author} {\bibfnamefont {S.}~\bibnamefont {Massar}},\ and\ \bibinfo
  {author} {\bibfnamefont {S.}~\bibnamefont {Popescu}},\ }\bibfield  {title}
  {\bibinfo {title} {Error filtration and entanglement purification for quantum
  communication},\ }\href {https://doi.org/10.1103/PhysRevA.72.012338}
  {\bibfield  {journal} {\bibinfo  {journal} {Phys. Rev. A}\ }\textbf {\bibinfo
  {volume} {72}},\ \bibinfo {pages} {012338} (\bibinfo {year}
  {2005})}\BibitemShut {NoStop}%
\bibitem [{\citenamefont {Lamoureux}\ \emph {et~al.}(2005)\citenamefont
  {Lamoureux}, \citenamefont {Brainis}, \citenamefont {Cerf}, \citenamefont
  {Emplit}, \citenamefont {Haelterman},\ and\ \citenamefont
  {Massar}}]{PhysRevLett.94.230501}%
  \BibitemOpen
  \bibfield  {author} {\bibinfo {author} {\bibfnamefont {L.-P.}\ \bibnamefont
  {Lamoureux}}, \bibinfo {author} {\bibfnamefont {E.}~\bibnamefont {Brainis}},
  \bibinfo {author} {\bibfnamefont {N.~J.}\ \bibnamefont {Cerf}}, \bibinfo
  {author} {\bibfnamefont {P.}~\bibnamefont {Emplit}}, \bibinfo {author}
  {\bibfnamefont {M.}~\bibnamefont {Haelterman}},\ and\ \bibinfo {author}
  {\bibfnamefont {S.}~\bibnamefont {Massar}},\ }\bibfield  {title} {\bibinfo
  {title} {Experimental error filtration for quantum communication over highly
  noisy channels},\ }\href {https://doi.org/10.1103/PhysRevLett.94.230501}
  {\bibfield  {journal} {\bibinfo  {journal} {Phys. Rev. Lett.}\ }\textbf
  {\bibinfo {volume} {94}},\ \bibinfo {pages} {230501} (\bibinfo {year}
  {2005})}\BibitemShut {NoStop}%
\bibitem [{\citenamefont {Vijayan}\ \emph {et~al.}(2020)\citenamefont
  {Vijayan}, \citenamefont {Lund},\ and\ \citenamefont
  {Rohde}}]{vijayan2020robust}%
  \BibitemOpen
  \bibfield  {author} {\bibinfo {author} {\bibfnamefont {M.~K.}\ \bibnamefont
  {Vijayan}}, \bibinfo {author} {\bibfnamefont {A.~P.}\ \bibnamefont {Lund}},\
  and\ \bibinfo {author} {\bibfnamefont {P.~P.}\ \bibnamefont {Rohde}},\
  }\bibfield  {title} {\bibinfo {title} {A robust w-state encoding for linear
  quantum optics},\ }\href@noop {} {\bibfield  {journal} {\bibinfo  {journal}
  {Quantum}\ }\textbf {\bibinfo {volume} {4}},\ \bibinfo {pages} {303}
  (\bibinfo {year} {2020})}\BibitemShut {NoStop}%
\bibitem [{\citenamefont {Lee}\ \emph {et~al.}(2022)\citenamefont {Lee},
  \citenamefont {Hann}, \citenamefont {Puri}, \citenamefont {Girvin},\ and\
  \citenamefont {Jiang}}]{lee2022error}%
  \BibitemOpen
  \bibfield  {author} {\bibinfo {author} {\bibfnamefont {G.}~\bibnamefont
  {Lee}}, \bibinfo {author} {\bibfnamefont {C.~T.}\ \bibnamefont {Hann}},
  \bibinfo {author} {\bibfnamefont {S.}~\bibnamefont {Puri}}, \bibinfo {author}
  {\bibfnamefont {S.}~\bibnamefont {Girvin}},\ and\ \bibinfo {author}
  {\bibfnamefont {L.}~\bibnamefont {Jiang}},\ }\bibfield  {title} {\bibinfo
  {title} {Error suppression for arbitrary-size black box quantum operations},\
  }\href@noop {} {\bibfield  {journal} {\bibinfo  {journal} {arXiv preprint
  arXiv:2210.10733}\ } (\bibinfo {year} {2022})}\BibitemShut {NoStop}%
\bibitem [{\citenamefont {Miguel-Ramiro}\ \emph
  {et~al.}(2023{\natexlab{a}})\citenamefont {Miguel-Ramiro}, \citenamefont
  {Shi}, \citenamefont {Dellantonio}, \citenamefont {Chan}, \citenamefont
  {Muschik},\ and\ \citenamefont {D{\"u}r}}]{miguel2023sqem}%
  \BibitemOpen
  \bibfield  {author} {\bibinfo {author} {\bibfnamefont {J.}~\bibnamefont
  {Miguel-Ramiro}}, \bibinfo {author} {\bibfnamefont {Z.}~\bibnamefont {Shi}},
  \bibinfo {author} {\bibfnamefont {L.}~\bibnamefont {Dellantonio}}, \bibinfo
  {author} {\bibfnamefont {A.}~\bibnamefont {Chan}}, \bibinfo {author}
  {\bibfnamefont {C.~A.}\ \bibnamefont {Muschik}},\ and\ \bibinfo {author}
  {\bibfnamefont {W.}~\bibnamefont {D{\"u}r}},\ }\bibfield  {title} {\bibinfo
  {title} {Sqem: Superposed quantum error mitigation},\ }\href@noop {}
  {\bibfield  {journal} {\bibinfo  {journal} {arXiv preprint arXiv:2304.08528}\
  } (\bibinfo {year} {2023}{\natexlab{a}})}\BibitemShut {NoStop}%
\bibitem [{\citenamefont {Miguel-Ramiro}\ \emph
  {et~al.}(2023{\natexlab{b}})\citenamefont {Miguel-Ramiro}, \citenamefont
  {Shi}, \citenamefont {Dellantonio}, \citenamefont {Chan}, \citenamefont
  {Muschik},\ and\ \citenamefont {D{\"u}r}}]{miguel2023enhancing}%
  \BibitemOpen
  \bibfield  {author} {\bibinfo {author} {\bibfnamefont {J.}~\bibnamefont
  {Miguel-Ramiro}}, \bibinfo {author} {\bibfnamefont {Z.}~\bibnamefont {Shi}},
  \bibinfo {author} {\bibfnamefont {L.}~\bibnamefont {Dellantonio}}, \bibinfo
  {author} {\bibfnamefont {A.}~\bibnamefont {Chan}}, \bibinfo {author}
  {\bibfnamefont {C.~A.}\ \bibnamefont {Muschik}},\ and\ \bibinfo {author}
  {\bibfnamefont {W.}~\bibnamefont {D{\"u}r}},\ }\bibfield  {title} {\bibinfo
  {title} {Enhancing quantum computation via superposition of quantum gates},\
  }\href@noop {} {\bibfield  {journal} {\bibinfo  {journal} {arXiv preprint
  arXiv:2304.08529}\ } (\bibinfo {year} {2023}{\natexlab{b}})}\BibitemShut
  {NoStop}%
\bibitem [{\citenamefont {Lim}\ and\ \citenamefont {Wang}(2021)}]{lim2021long}%
  \BibitemOpen
  \bibfield  {author} {\bibinfo {author} {\bibfnamefont {C.~C.-W.}\
  \bibnamefont {Lim}}\ and\ \bibinfo {author} {\bibfnamefont {C.}~\bibnamefont
  {Wang}},\ }\bibfield  {title} {\bibinfo {title} {Long-distance quantum key
  distribution gets real},\ }\href@noop {} {\bibfield  {journal} {\bibinfo
  {journal} {Nature Photonics}\ }\textbf {\bibinfo {volume} {15}},\ \bibinfo
  {pages} {554} (\bibinfo {year} {2021})}\BibitemShut {NoStop}%
\bibitem [{\citenamefont {Arvidsson-Shukur}\ \emph {et~al.}(2020)\citenamefont
  {Arvidsson-Shukur}, \citenamefont {Halpern}, \citenamefont {Lepage},
  \citenamefont {Lasek}, \citenamefont {Barnes},\ and\ \citenamefont
  {Lloyd}}]{postmetro}%
  \BibitemOpen
  \bibfield  {author} {\bibinfo {author} {\bibfnamefont {D.~R.~M.}\
  \bibnamefont {Arvidsson-Shukur}}, \bibinfo {author} {\bibfnamefont {N.~Y.}\
  \bibnamefont {Halpern}}, \bibinfo {author} {\bibfnamefont {H.~V.}\
  \bibnamefont {Lepage}}, \bibinfo {author} {\bibfnamefont {A.~A.}\
  \bibnamefont {Lasek}}, \bibinfo {author} {\bibfnamefont {C.~H.~W.}\
  \bibnamefont {Barnes}},\ and\ \bibinfo {author} {\bibfnamefont
  {S.}~\bibnamefont {Lloyd}},\ }\bibfield  {title} {\bibinfo {title} {Quantum
  advantage in postselected metrology},\ }\href
  {https://doi.org/10.1038/s41467-020-17559-w} {\bibfield  {journal} {\bibinfo
  {journal} {Nature Communications}\ }\textbf {\bibinfo {volume} {11}},\
  \bibinfo {pages} {3775} (\bibinfo {year} {2020})}\BibitemShut {NoStop}%
\bibitem [{\citenamefont {Lami}\ and\ \citenamefont
  {Wilde}(2023)}]{lami2023exact}%
  \BibitemOpen
  \bibfield  {author} {\bibinfo {author} {\bibfnamefont {L.}~\bibnamefont
  {Lami}}\ and\ \bibinfo {author} {\bibfnamefont {M.~M.}\ \bibnamefont
  {Wilde}},\ }\bibfield  {title} {\bibinfo {title} {Exact solution for the
  quantum and private capacities of bosonic dephasing channels},\ }\href@noop
  {} {\bibfield  {journal} {\bibinfo  {journal} {Nature Photonics}\ }\textbf
  {\bibinfo {volume} {17}},\ \bibinfo {pages} {525} (\bibinfo {year}
  {2023})}\BibitemShut {NoStop}%
\bibitem [{\citenamefont {Marchese}\ and\ \citenamefont
  {Kok}(2023)}]{PhysRevLett.130.160801}%
  \BibitemOpen
  \bibfield  {author} {\bibinfo {author} {\bibfnamefont {M.~M.}\ \bibnamefont
  {Marchese}}\ and\ \bibinfo {author} {\bibfnamefont {P.}~\bibnamefont {Kok}},\
  }\bibfield  {title} {\bibinfo {title} {Large baseline optical imaging
  assisted by single photons and linear quantum optics},\ }\href
  {https://doi.org/10.1103/PhysRevLett.130.160801} {\bibfield  {journal}
  {\bibinfo  {journal} {Phys. Rev. Lett.}\ }\textbf {\bibinfo {volume} {130}},\
  \bibinfo {pages} {160801} (\bibinfo {year} {2023})}\BibitemShut {NoStop}%
\bibitem [{\citenamefont {Khabiboulline}\ \emph
  {et~al.}(2019{\natexlab{a}})\citenamefont {Khabiboulline}, \citenamefont
  {Borregaard}, \citenamefont {De~Greve},\ and\ \citenamefont
  {Lukin}}]{PhysRevLett.123.070504}%
  \BibitemOpen
  \bibfield  {author} {\bibinfo {author} {\bibfnamefont {E.~T.}\ \bibnamefont
  {Khabiboulline}}, \bibinfo {author} {\bibfnamefont {J.}~\bibnamefont
  {Borregaard}}, \bibinfo {author} {\bibfnamefont {K.}~\bibnamefont
  {De~Greve}},\ and\ \bibinfo {author} {\bibfnamefont {M.~D.}\ \bibnamefont
  {Lukin}},\ }\bibfield  {title} {\bibinfo {title} {Optical interferometry with
  quantum networks},\ }\href {https://doi.org/10.1103/PhysRevLett.123.070504}
  {\bibfield  {journal} {\bibinfo  {journal} {Phys. Rev. Lett.}\ }\textbf
  {\bibinfo {volume} {123}},\ \bibinfo {pages} {070504} (\bibinfo {year}
  {2019}{\natexlab{a}})}\BibitemShut {NoStop}%
\bibitem [{\citenamefont {Khabiboulline}\ \emph
  {et~al.}(2019{\natexlab{b}})\citenamefont {Khabiboulline}, \citenamefont
  {Borregaard}, \citenamefont {De~Greve},\ and\ \citenamefont
  {Lukin}}]{PhysRevA.100.022316}%
  \BibitemOpen
  \bibfield  {author} {\bibinfo {author} {\bibfnamefont {E.~T.}\ \bibnamefont
  {Khabiboulline}}, \bibinfo {author} {\bibfnamefont {J.}~\bibnamefont
  {Borregaard}}, \bibinfo {author} {\bibfnamefont {K.}~\bibnamefont
  {De~Greve}},\ and\ \bibinfo {author} {\bibfnamefont {M.~D.}\ \bibnamefont
  {Lukin}},\ }\bibfield  {title} {\bibinfo {title} {Quantum-assisted telescope
  arrays},\ }\href {https://doi.org/10.1103/PhysRevA.100.022316} {\bibfield
  {journal} {\bibinfo  {journal} {Phys. Rev. A}\ }\textbf {\bibinfo {volume}
  {100}},\ \bibinfo {pages} {022316} (\bibinfo {year}
  {2019}{\natexlab{b}})}\BibitemShut {NoStop}%
\bibitem [{\citenamefont {Huang}\ \emph {et~al.}(2022)\citenamefont {Huang},
  \citenamefont {Brennen},\ and\ \citenamefont
  {Ouyang}}]{PhysRevLett.129.210502}%
  \BibitemOpen
  \bibfield  {author} {\bibinfo {author} {\bibfnamefont {Z.}~\bibnamefont
  {Huang}}, \bibinfo {author} {\bibfnamefont {G.~K.}\ \bibnamefont {Brennen}},\
  and\ \bibinfo {author} {\bibfnamefont {Y.}~\bibnamefont {Ouyang}},\
  }\bibfield  {title} {\bibinfo {title} {Imaging stars with quantum error
  correction},\ }\href {https://doi.org/10.1103/PhysRevLett.129.210502}
  {\bibfield  {journal} {\bibinfo  {journal} {Phys. Rev. Lett.}\ }\textbf
  {\bibinfo {volume} {129}},\ \bibinfo {pages} {210502} (\bibinfo {year}
  {2022})}\BibitemShut {NoStop}%
\bibitem [{\citenamefont {McAlister}\ \emph {et~al.}(2005)\citenamefont
  {McAlister}, \citenamefont {ten Brummelaar}, \citenamefont {Gies},
  \citenamefont {Huang}, \citenamefont {W.~G.~Bagnuolo}, \citenamefont {Shure},
  \citenamefont {Sturmann}, \citenamefont {Sturmann}, \citenamefont {Turner},
  \citenamefont {Taylor}, \citenamefont {Berger}, \citenamefont {Baines},
  \citenamefont {Grundstrom}, \citenamefont {Ogden}, \citenamefont {Ridgway},\
  and\ \citenamefont {van Belle}}]{McAlister_2005}%
  \BibitemOpen
  \bibfield  {author} {\bibinfo {author} {\bibfnamefont {H.~A.}\ \bibnamefont
  {McAlister}}, \bibinfo {author} {\bibfnamefont {T.~A.}\ \bibnamefont {ten
  Brummelaar}}, \bibinfo {author} {\bibfnamefont {D.~R.}\ \bibnamefont {Gies}},
  \bibinfo {author} {\bibfnamefont {W.}~\bibnamefont {Huang}}, \bibinfo
  {author} {\bibfnamefont {J.}~\bibnamefont {W.~G.~Bagnuolo}}, \bibinfo
  {author} {\bibfnamefont {M.~A.}\ \bibnamefont {Shure}}, \bibinfo {author}
  {\bibfnamefont {J.}~\bibnamefont {Sturmann}}, \bibinfo {author}
  {\bibfnamefont {L.}~\bibnamefont {Sturmann}}, \bibinfo {author}
  {\bibfnamefont {N.~H.}\ \bibnamefont {Turner}}, \bibinfo {author}
  {\bibfnamefont {S.~F.}\ \bibnamefont {Taylor}}, \bibinfo {author}
  {\bibfnamefont {D.~H.}\ \bibnamefont {Berger}}, \bibinfo {author}
  {\bibfnamefont {E.~K.}\ \bibnamefont {Baines}}, \bibinfo {author}
  {\bibfnamefont {E.}~\bibnamefont {Grundstrom}}, \bibinfo {author}
  {\bibfnamefont {C.}~\bibnamefont {Ogden}}, \bibinfo {author} {\bibfnamefont
  {S.~T.}\ \bibnamefont {Ridgway}},\ and\ \bibinfo {author} {\bibfnamefont
  {G.}~\bibnamefont {van Belle}},\ }\bibfield  {title} {\bibinfo {title} {First
  results from the chara array. i. an interferometric and spectroscopic study
  of the fast rotator $\alpha$ leonis (regulus)},\ }\href
  {https://doi.org/10.1086/430730} {\bibfield  {journal} {\bibinfo  {journal}
  {The Astrophysical Journal}\ }\textbf {\bibinfo {volume} {628}},\ \bibinfo
  {pages} {439} (\bibinfo {year} {2005})}\BibitemShut {NoStop}%
\bibitem [{\citenamefont {Anugu}\ \emph {et~al.}(2020)\citenamefont {Anugu},
  \citenamefont {Bouquin}, \citenamefont {Monnier}, \citenamefont {Kraus},
  \citenamefont {Setterholm}, \citenamefont {Labdon}, \citenamefont {Davies},
  \citenamefont {Lanthermann}, \citenamefont {Gardner}, \citenamefont {Ennis},
  \citenamefont {Johnson}, \citenamefont {Brummelaar}, \citenamefont
  {Schaefer},\ and\ \citenamefont {Sturmann}}]{Anugu_2020}%
  \BibitemOpen
  \bibfield  {author} {\bibinfo {author} {\bibfnamefont {N.}~\bibnamefont
  {Anugu}}, \bibinfo {author} {\bibfnamefont {J.-B.~L.}\ \bibnamefont
  {Bouquin}}, \bibinfo {author} {\bibfnamefont {J.~D.}\ \bibnamefont
  {Monnier}}, \bibinfo {author} {\bibfnamefont {S.}~\bibnamefont {Kraus}},
  \bibinfo {author} {\bibfnamefont {B.~R.}\ \bibnamefont {Setterholm}},
  \bibinfo {author} {\bibfnamefont {A.}~\bibnamefont {Labdon}}, \bibinfo
  {author} {\bibfnamefont {C.~L.}\ \bibnamefont {Davies}}, \bibinfo {author}
  {\bibfnamefont {C.}~\bibnamefont {Lanthermann}}, \bibinfo {author}
  {\bibfnamefont {T.}~\bibnamefont {Gardner}}, \bibinfo {author} {\bibfnamefont
  {J.}~\bibnamefont {Ennis}}, \bibinfo {author} {\bibfnamefont {K.~J.~C.}\
  \bibnamefont {Johnson}}, \bibinfo {author} {\bibfnamefont {T.~T.}\
  \bibnamefont {Brummelaar}}, \bibinfo {author} {\bibfnamefont
  {G.}~\bibnamefont {Schaefer}},\ and\ \bibinfo {author} {\bibfnamefont
  {J.}~\bibnamefont {Sturmann}},\ }\bibfield  {title} {\bibinfo {title}
  {Mirc-x: A highly sensitive six-telescope interferometric imager at the chara
  array},\ }\href {https://doi.org/10.3847/1538-3881/aba957} {\bibfield
  {journal} {\bibinfo  {journal} {The Astronomical Journal}\ }\textbf {\bibinfo
  {volume} {160}},\ \bibinfo {pages} {158} (\bibinfo {year}
  {2020})}\BibitemShut {NoStop}%
\bibitem [{\citenamefont {Wehner}\ \emph {et~al.}(2018)\citenamefont {Wehner},
  \citenamefont {Elkouss},\ and\ \citenamefont
  {Hanson}}]{doi:10.1126/science.aam9288}%
  \BibitemOpen
  \bibfield  {author} {\bibinfo {author} {\bibfnamefont {S.}~\bibnamefont
  {Wehner}}, \bibinfo {author} {\bibfnamefont {D.}~\bibnamefont {Elkouss}},\
  and\ \bibinfo {author} {\bibfnamefont {R.}~\bibnamefont {Hanson}},\
  }\bibfield  {title} {\bibinfo {title} {Quantum internet: A vision for the
  road ahead},\ }\href {https://doi.org/10.1126/science.aam9288} {\bibfield
  {journal} {\bibinfo  {journal} {Science}\ }\textbf {\bibinfo {volume}
  {362}},\ \bibinfo {pages} {eaam9288} (\bibinfo {year} {2018})},\ \Eprint
  {https://arxiv.org/abs/https://www.science.org/doi/pdf/10.1126/science.aam9288}
  {https://www.science.org/doi/pdf/10.1126/science.aam9288} \BibitemShut
  {NoStop}%
\bibitem [{\citenamefont {Kok}\ \emph {et~al.}(2007)\citenamefont {Kok},
  \citenamefont {Munro}, \citenamefont {Nemoto}, \citenamefont {Ralph},
  \citenamefont {Dowling},\ and\ \citenamefont {Milburn}}]{RevModPhys.79.135}%
  \BibitemOpen
  \bibfield  {author} {\bibinfo {author} {\bibfnamefont {P.}~\bibnamefont
  {Kok}}, \bibinfo {author} {\bibfnamefont {W.~J.}\ \bibnamefont {Munro}},
  \bibinfo {author} {\bibfnamefont {K.}~\bibnamefont {Nemoto}}, \bibinfo
  {author} {\bibfnamefont {T.~C.}\ \bibnamefont {Ralph}}, \bibinfo {author}
  {\bibfnamefont {J.~P.}\ \bibnamefont {Dowling}},\ and\ \bibinfo {author}
  {\bibfnamefont {G.~J.}\ \bibnamefont {Milburn}},\ }\bibfield  {title}
  {\bibinfo {title} {Linear optical quantum computing with photonic qubits},\
  }\href {https://doi.org/10.1103/RevModPhys.79.135} {\bibfield  {journal}
  {\bibinfo  {journal} {Rev. Mod. Phys.}\ }\textbf {\bibinfo {volume} {79}},\
  \bibinfo {pages} {135} (\bibinfo {year} {2007})}\BibitemShut {NoStop}%
\bibitem [{\citenamefont {Tan}\ \emph {et~al.}(1991)\citenamefont {Tan},
  \citenamefont {Walls},\ and\ \citenamefont {Collett}}]{PhysRevLett.66.252}%
  \BibitemOpen
  \bibfield  {author} {\bibinfo {author} {\bibfnamefont {S.~M.}\ \bibnamefont
  {Tan}}, \bibinfo {author} {\bibfnamefont {D.~F.}\ \bibnamefont {Walls}},\
  and\ \bibinfo {author} {\bibfnamefont {M.~J.}\ \bibnamefont {Collett}},\
  }\bibfield  {title} {\bibinfo {title} {Nonlocality of a single photon},\
  }\href {https://doi.org/10.1103/PhysRevLett.66.252} {\bibfield  {journal}
  {\bibinfo  {journal} {Phys. Rev. Lett.}\ }\textbf {\bibinfo {volume} {66}},\
  \bibinfo {pages} {252} (\bibinfo {year} {1991})}\BibitemShut {NoStop}%
\bibitem [{\citenamefont {Reck}\ \emph {et~al.}(1994)\citenamefont {Reck},
  \citenamefont {Zeilinger}, \citenamefont {Bernstein},\ and\ \citenamefont
  {Bertani}}]{PhysRevLett.73.58}%
  \BibitemOpen
  \bibfield  {author} {\bibinfo {author} {\bibfnamefont {M.}~\bibnamefont
  {Reck}}, \bibinfo {author} {\bibfnamefont {A.}~\bibnamefont {Zeilinger}},
  \bibinfo {author} {\bibfnamefont {H.~J.}\ \bibnamefont {Bernstein}},\ and\
  \bibinfo {author} {\bibfnamefont {P.}~\bibnamefont {Bertani}},\ }\bibfield
  {title} {\bibinfo {title} {Experimental realization of any discrete unitary
  operator},\ }\href {https://doi.org/10.1103/PhysRevLett.73.58} {\bibfield
  {journal} {\bibinfo  {journal} {Phys. Rev. Lett.}\ }\textbf {\bibinfo
  {volume} {73}},\ \bibinfo {pages} {58} (\bibinfo {year} {1994})}\BibitemShut
  {NoStop}%
\bibitem [{\citenamefont {Ip}\ \emph {et~al.}(2022)\citenamefont {Ip},
  \citenamefont {Ravet}, \citenamefont {Martins}, \citenamefont {Huang},
  \citenamefont {Okamoto}, \citenamefont {Han}, \citenamefont {Narisetty},
  \citenamefont {Fang}, \citenamefont {Huang}, \citenamefont {Salemi} \emph
  {et~al.}}]{ip2022using}%
  \BibitemOpen
  \bibfield  {author} {\bibinfo {author} {\bibfnamefont {E.}~\bibnamefont
  {Ip}}, \bibinfo {author} {\bibfnamefont {F.}~\bibnamefont {Ravet}}, \bibinfo
  {author} {\bibfnamefont {H.}~\bibnamefont {Martins}}, \bibinfo {author}
  {\bibfnamefont {M.-F.}\ \bibnamefont {Huang}}, \bibinfo {author}
  {\bibfnamefont {T.}~\bibnamefont {Okamoto}}, \bibinfo {author} {\bibfnamefont
  {S.}~\bibnamefont {Han}}, \bibinfo {author} {\bibfnamefont {C.}~\bibnamefont
  {Narisetty}}, \bibinfo {author} {\bibfnamefont {J.}~\bibnamefont {Fang}},
  \bibinfo {author} {\bibfnamefont {Y.-K.}\ \bibnamefont {Huang}}, \bibinfo
  {author} {\bibfnamefont {M.}~\bibnamefont {Salemi}}, \emph {et~al.},\
  }\bibfield  {title} {\bibinfo {title} {Using global existing fiber networks
  for environmental sensing},\ }\href@noop {} {\bibfield  {journal} {\bibinfo
  {journal} {Proceedings of the IEEE}\ }\textbf {\bibinfo {volume} {110}},\
  \bibinfo {pages} {1853} (\bibinfo {year} {2022})}\BibitemShut {NoStop}%
\bibitem [{\citenamefont {Hsieh}\ and\ \citenamefont
  {Hung}(1996)}]{hsieh1996phase}%
  \BibitemOpen
  \bibfield  {author} {\bibinfo {author} {\bibfnamefont {G.-C.}\ \bibnamefont
  {Hsieh}}\ and\ \bibinfo {author} {\bibfnamefont {J.~C.}\ \bibnamefont
  {Hung}},\ }\bibfield  {title} {\bibinfo {title} {Phase-locked loop
  techniques. a survey},\ }\href@noop {} {\bibfield  {journal} {\bibinfo
  {journal} {IEEE Transactions on industrial electronics}\ }\textbf {\bibinfo
  {volume} {43}},\ \bibinfo {pages} {609} (\bibinfo {year} {1996})}\BibitemShut
  {NoStop}%
\bibitem [{\citenamefont {Minder}\ \emph {et~al.}(2019)\citenamefont {Minder},
  \citenamefont {Pittaluga}, \citenamefont {Roberts}, \citenamefont
  {Lucamarini}, \citenamefont {Dynes}, \citenamefont {Yuan},\ and\
  \citenamefont {Shields}}]{minder2019experimental}%
  \BibitemOpen
  \bibfield  {author} {\bibinfo {author} {\bibfnamefont {M.}~\bibnamefont
  {Minder}}, \bibinfo {author} {\bibfnamefont {M.}~\bibnamefont {Pittaluga}},
  \bibinfo {author} {\bibfnamefont {G.~L.}\ \bibnamefont {Roberts}}, \bibinfo
  {author} {\bibfnamefont {M.}~\bibnamefont {Lucamarini}}, \bibinfo {author}
  {\bibfnamefont {J.~F.}\ \bibnamefont {Dynes}}, \bibinfo {author}
  {\bibfnamefont {Z.}~\bibnamefont {Yuan}},\ and\ \bibinfo {author}
  {\bibfnamefont {A.~J.}\ \bibnamefont {Shields}},\ }\bibfield  {title}
  {\bibinfo {title} {Experimental quantum key distribution beyond the
  repeaterless secret key capacity},\ }\href@noop {} {\bibfield  {journal}
  {\bibinfo  {journal} {Nature Photonics}\ }\textbf {\bibinfo {volume} {13}},\
  \bibinfo {pages} {334} (\bibinfo {year} {2019})}\BibitemShut {NoStop}%
\bibitem [{\citenamefont {Grosshans}\ \emph {et~al.}(2003)\citenamefont
  {Grosshans}, \citenamefont {Van~Assche}, \citenamefont {Wenger},
  \citenamefont {Brouri}, \citenamefont {Cerf},\ and\ \citenamefont
  {Grangier}}]{grosshans2003quantum}%
  \BibitemOpen
  \bibfield  {author} {\bibinfo {author} {\bibfnamefont {F.}~\bibnamefont
  {Grosshans}}, \bibinfo {author} {\bibfnamefont {G.}~\bibnamefont
  {Van~Assche}}, \bibinfo {author} {\bibfnamefont {J.}~\bibnamefont {Wenger}},
  \bibinfo {author} {\bibfnamefont {R.}~\bibnamefont {Brouri}}, \bibinfo
  {author} {\bibfnamefont {N.~J.}\ \bibnamefont {Cerf}},\ and\ \bibinfo
  {author} {\bibfnamefont {P.}~\bibnamefont {Grangier}},\ }\bibfield  {title}
  {\bibinfo {title} {Quantum key distribution using gaussian-modulated coherent
  states},\ }\href@noop {} {\bibfield  {journal} {\bibinfo  {journal} {Nature}\
  }\textbf {\bibinfo {volume} {421}},\ \bibinfo {pages} {238} (\bibinfo {year}
  {2003})}\BibitemShut {NoStop}%
\bibitem [{\citenamefont {Wang}\ \emph {et~al.}(2022)\citenamefont {Wang},
  \citenamefont {Yin}, \citenamefont {He}, \citenamefont {Chen}, \citenamefont
  {Wang}, \citenamefont {Ye}, \citenamefont {Zhou}, \citenamefont {Fan-Yuan},
  \citenamefont {Wang}, \citenamefont {Chen} \emph {et~al.}}]{wang2022twin}%
  \BibitemOpen
  \bibfield  {author} {\bibinfo {author} {\bibfnamefont {S.}~\bibnamefont
  {Wang}}, \bibinfo {author} {\bibfnamefont {Z.-Q.}\ \bibnamefont {Yin}},
  \bibinfo {author} {\bibfnamefont {D.-Y.}\ \bibnamefont {He}}, \bibinfo
  {author} {\bibfnamefont {W.}~\bibnamefont {Chen}}, \bibinfo {author}
  {\bibfnamefont {R.-Q.}\ \bibnamefont {Wang}}, \bibinfo {author}
  {\bibfnamefont {P.}~\bibnamefont {Ye}}, \bibinfo {author} {\bibfnamefont
  {Y.}~\bibnamefont {Zhou}}, \bibinfo {author} {\bibfnamefont {G.-J.}\
  \bibnamefont {Fan-Yuan}}, \bibinfo {author} {\bibfnamefont {F.-X.}\
  \bibnamefont {Wang}}, \bibinfo {author} {\bibfnamefont {W.}~\bibnamefont
  {Chen}}, \emph {et~al.},\ }\bibfield  {title} {\bibinfo {title} {Twin-field
  quantum key distribution over 830-km fibre},\ }\href@noop {} {\bibfield
  {journal} {\bibinfo  {journal} {Nature photonics}\ }\textbf {\bibinfo
  {volume} {16}},\ \bibinfo {pages} {154} (\bibinfo {year} {2022})}\BibitemShut
  {NoStop}%
\bibitem [{\citenamefont {Mandel}\ and\ \citenamefont
  {Wolf}(1995)}]{mandel1995optical}%
  \BibitemOpen
  \bibfield  {author} {\bibinfo {author} {\bibfnamefont {L.}~\bibnamefont
  {Mandel}}\ and\ \bibinfo {author} {\bibfnamefont {E.}~\bibnamefont {Wolf}},\
  }\href@noop {} {\emph {\bibinfo {title} {Optical coherence and quantum
  optics}}}\ (\bibinfo  {publisher} {Cambridge university press},\ \bibinfo
  {year} {1995})\BibitemShut {NoStop}%
\bibitem [{\citenamefont {Tsang}(2011)}]{PhysRevLett.107.270402}%
  \BibitemOpen
  \bibfield  {author} {\bibinfo {author} {\bibfnamefont {M.}~\bibnamefont
  {Tsang}},\ }\bibfield  {title} {\bibinfo {title} {Quantum nonlocality in
  weak-thermal-light interferometry},\ }\href
  {https://doi.org/10.1103/PhysRevLett.107.270402} {\bibfield  {journal}
  {\bibinfo  {journal} {Phys. Rev. Lett.}\ }\textbf {\bibinfo {volume} {107}},\
  \bibinfo {pages} {270402} (\bibinfo {year} {2011})}\BibitemShut {NoStop}%
\bibitem [{\citenamefont {Braunstein}\ and\ \citenamefont
  {Caves}(1994)}]{caves}%
  \BibitemOpen
  \bibfield  {author} {\bibinfo {author} {\bibfnamefont {S.~L.}\ \bibnamefont
  {Braunstein}}\ and\ \bibinfo {author} {\bibfnamefont {C.~M.}\ \bibnamefont
  {Caves}},\ }\bibfield  {title} {\bibinfo {title} {Statistical distance and
  the geometry of quantum states},\ }\href
  {https://doi.org/10.1103/PhysRevLett.72.3439} {\bibfield  {journal} {\bibinfo
   {journal} {Physical Review Letters}\ }\textbf {\bibinfo {volume} {72}},\
  \bibinfo {pages} {3439} (\bibinfo {year} {1994})}\BibitemShut {NoStop}%
\bibitem [{\citenamefont {Braunstein}\ \emph {et~al.}(1996)\citenamefont
  {Braunstein}, \citenamefont {Caves},\ and\ \citenamefont {Milburn}}]{caves1}%
  \BibitemOpen
  \bibfield  {author} {\bibinfo {author} {\bibfnamefont {S.~L.}\ \bibnamefont
  {Braunstein}}, \bibinfo {author} {\bibfnamefont {C.~M.}\ \bibnamefont
  {Caves}},\ and\ \bibinfo {author} {\bibfnamefont {G.~J.}\ \bibnamefont
  {Milburn}},\ }\bibfield  {title} {\bibinfo {title} {Generalized uncertainty
  relations: Theory, examples, and {L}orentz invariance},\ }\href@noop {}
  {\bibfield  {journal} {\bibinfo  {journal} {Annals of Physics}\ }\textbf
  {\bibinfo {volume} {247}},\ \bibinfo {pages} {135} (\bibinfo {year}
  {1996})}\BibitemShut {NoStop}%
\bibitem [{\citenamefont {Giovannetti}\ \emph {et~al.}(2011)\citenamefont
  {Giovannetti}, \citenamefont {Lloyd},\ and\ \citenamefont
  {Maccone}}]{giovannetti2011advances}%
  \BibitemOpen
  \bibfield  {author} {\bibinfo {author} {\bibfnamefont {V.}~\bibnamefont
  {Giovannetti}}, \bibinfo {author} {\bibfnamefont {S.}~\bibnamefont {Lloyd}},\
  and\ \bibinfo {author} {\bibfnamefont {L.}~\bibnamefont {Maccone}},\
  }\bibfield  {title} {\bibinfo {title} {Advances in quantum metrology},\
  }\href {https://doi.org/10.1038/nphoton.2011.35} {\bibfield  {journal}
  {\bibinfo  {journal} {Nature Photonics}\ }\textbf {\bibinfo {volume} {5}},\
  \bibinfo {pages} {222} (\bibinfo {year} {2011})}\BibitemShut {NoStop}%
\bibitem [{\citenamefont {Giovannetti}\ \emph {et~al.}(2006)\citenamefont
  {Giovannetti}, \citenamefont {Lloyd},\ and\ \citenamefont
  {Maccone}}]{giovannetti2006quantum}%
  \BibitemOpen
  \bibfield  {author} {\bibinfo {author} {\bibfnamefont {V.}~\bibnamefont
  {Giovannetti}}, \bibinfo {author} {\bibfnamefont {S.}~\bibnamefont {Lloyd}},\
  and\ \bibinfo {author} {\bibfnamefont {L.}~\bibnamefont {Maccone}},\
  }\bibfield  {title} {\bibinfo {title} {Quantum metrology},\ }\href
  {https://doi.org/10.1103/PhysRevLett.96.010401} {\bibfield  {journal}
  {\bibinfo  {journal} {Physical Review Letters}\ }\textbf {\bibinfo {volume}
  {96}},\ \bibinfo {pages} {010401} (\bibinfo {year} {2006})}\BibitemShut
  {NoStop}%
\bibitem [{\citenamefont {Paris}(2009{\natexlab{a}})}]{paris2009quantum}%
  \BibitemOpen
  \bibfield  {author} {\bibinfo {author} {\bibfnamefont {M.~G.}\ \bibnamefont
  {Paris}},\ }\bibfield  {title} {\bibinfo {title} {Quantum estimation for
  quantum technology},\ }\href@noop {} {\bibfield  {journal} {\bibinfo
  {journal} {International Journal of Quantum Information}\ }\textbf {\bibinfo
  {volume} {7}},\ \bibinfo {pages} {125} (\bibinfo {year}
  {2009}{\natexlab{a}})}\BibitemShut {NoStop}%
\bibitem [{\citenamefont {Barndorff-Nielsen}\ and\ \citenamefont
  {Gill}(2000)}]{barndorff2000fisher}%
  \BibitemOpen
  \bibfield  {author} {\bibinfo {author} {\bibfnamefont {O.~E.}\ \bibnamefont
  {Barndorff-Nielsen}}\ and\ \bibinfo {author} {\bibfnamefont {R.~D.}\
  \bibnamefont {Gill}},\ }\bibfield  {title} {\bibinfo {title} {Fisher
  information in quantum statistics},\ }\href@noop {} {\bibfield  {journal}
  {\bibinfo  {journal} {Journal of Physics A: Mathematical and General}\
  }\textbf {\bibinfo {volume} {33}},\ \bibinfo {pages} {4481} (\bibinfo {year}
  {2000})}\BibitemShut {NoStop}%
\bibitem [{\citenamefont {Paris}(2009{\natexlab{b}})}]{Paris}%
  \BibitemOpen
  \bibfield  {author} {\bibinfo {author} {\bibfnamefont {M.~G.~A.}\
  \bibnamefont {Paris}},\ }\bibfield  {title} {\bibinfo {title} {Quantum
  estimation for quantum technology},\ }\href
  {https://doi.org/10.1142/S0219749909004839} {\bibfield  {journal} {\bibinfo
  {journal} {International Journal of Quantum Information}\ }\textbf {\bibinfo
  {volume} {07}},\ \bibinfo {pages} {125} (\bibinfo {year}
  {2009}{\natexlab{b}})},\ \Eprint
  {https://arxiv.org/abs/https://doi.org/10.1142/S0219749909004839}
  {https://doi.org/10.1142/S0219749909004839} \BibitemShut {NoStop}%
\bibitem [{\citenamefont {Sidhu}\ and\ \citenamefont {Kok}(2020)}]{Sidhu}%
  \BibitemOpen
  \bibfield  {author} {\bibinfo {author} {\bibfnamefont {J.~S.}\ \bibnamefont
  {Sidhu}}\ and\ \bibinfo {author} {\bibfnamefont {P.}~\bibnamefont {Kok}},\
  }\bibfield  {title} {\bibinfo {title} {{Geometric perspective on quantum
  parameter estimation}},\ }\href {https://doi.org/10.1116/1.5119961}
  {\bibfield  {journal} {\bibinfo  {journal} {AVS Quantum Science}\ }\textbf
  {\bibinfo {volume} {2}},\ \bibinfo {pages} {014701} (\bibinfo {year}
  {2020})},\ \Eprint
  {https://arxiv.org/abs/https://pubs.aip.org/avs/aqs/article-pdf/doi/10.1116/1.5119961/16700179/014701\_1\_online.pdf}
  {https://pubs.aip.org/avs/aqs/article-pdf/doi/10.1116/1.5119961/16700179/014701\_1\_online.pdf}
  \BibitemShut {NoStop}%
\bibitem [{\citenamefont {Pearce}\ \emph {et~al.}(2017)\citenamefont {Pearce},
  \citenamefont {Campbell},\ and\ \citenamefont {Kok}}]{Pearce2017optimal}%
  \BibitemOpen
  \bibfield  {author} {\bibinfo {author} {\bibfnamefont {M.~E.}\ \bibnamefont
  {Pearce}}, \bibinfo {author} {\bibfnamefont {E.~T.}\ \bibnamefont
  {Campbell}},\ and\ \bibinfo {author} {\bibfnamefont {P.}~\bibnamefont
  {Kok}},\ }\bibfield  {title} {\bibinfo {title} {Optimal quantum metrology of
  distant black bodies},\ }\href@noop {} {\bibfield  {journal} {\bibinfo
  {journal} {Quantum}\ }\textbf {\bibinfo {volume} {1}},\ \bibinfo {pages} {21}
  (\bibinfo {year} {2017})}\BibitemShut {NoStop}%
\bibitem [{\citenamefont {Huang}\ \emph {et~al.}(2024)\citenamefont {Huang},
  \citenamefont {Baragiola}, \citenamefont {Menicucci},\ and\ \citenamefont
  {Wilde}}]{PhysRevA.109.052434}%
  \BibitemOpen
  \bibfield  {author} {\bibinfo {author} {\bibfnamefont {Z.}~\bibnamefont
  {Huang}}, \bibinfo {author} {\bibfnamefont {B.~Q.}\ \bibnamefont
  {Baragiola}}, \bibinfo {author} {\bibfnamefont {N.~C.}\ \bibnamefont
  {Menicucci}},\ and\ \bibinfo {author} {\bibfnamefont {M.~M.}\ \bibnamefont
  {Wilde}},\ }\bibfield  {title} {\bibinfo {title} {Limited quantum advantage
  for stellar interferometry via continuous-variable teleportation},\ }\href
  {https://doi.org/10.1103/PhysRevA.109.052434} {\bibfield  {journal} {\bibinfo
   {journal} {Phys. Rev. A}\ }\textbf {\bibinfo {volume} {109}},\ \bibinfo
  {pages} {052434} (\bibinfo {year} {2024})}\BibitemShut {NoStop}%
\bibitem [{\citenamefont {Wang}\ \emph
  {et~al.}(2019{\natexlab{a}})\citenamefont {Wang}, \citenamefont {He},
  \citenamefont {Yin}, \citenamefont {Lu}, \citenamefont {Cui}, \citenamefont
  {Chen}, \citenamefont {Zhou}, \citenamefont {Guo},\ and\ \citenamefont
  {Han}}]{PhysRevX.9.021046}%
  \BibitemOpen
  \bibfield  {author} {\bibinfo {author} {\bibfnamefont {S.}~\bibnamefont
  {Wang}}, \bibinfo {author} {\bibfnamefont {D.-Y.}\ \bibnamefont {He}},
  \bibinfo {author} {\bibfnamefont {Z.-Q.}\ \bibnamefont {Yin}}, \bibinfo
  {author} {\bibfnamefont {F.-Y.}\ \bibnamefont {Lu}}, \bibinfo {author}
  {\bibfnamefont {C.-H.}\ \bibnamefont {Cui}}, \bibinfo {author} {\bibfnamefont
  {W.}~\bibnamefont {Chen}}, \bibinfo {author} {\bibfnamefont {Z.}~\bibnamefont
  {Zhou}}, \bibinfo {author} {\bibfnamefont {G.-C.}\ \bibnamefont {Guo}},\ and\
  \bibinfo {author} {\bibfnamefont {Z.-F.}\ \bibnamefont {Han}},\ }\bibfield
  {title} {\bibinfo {title} {Beating the fundamental rate-distance limit in a
  proof-of-principle quantum key distribution system},\ }\href
  {https://doi.org/10.1103/PhysRevX.9.021046} {\bibfield  {journal} {\bibinfo
  {journal} {Phys. Rev. X}\ }\textbf {\bibinfo {volume} {9}},\ \bibinfo {pages}
  {021046} (\bibinfo {year} {2019}{\natexlab{a}})}\BibitemShut {NoStop}%
\bibitem [{\citenamefont {Amies-King}\ \emph {et~al.}(2023)\citenamefont
  {Amies-King}, \citenamefont {Schatz}, \citenamefont {Duan}, \citenamefont
  {Biswas}, \citenamefont {Bailey}, \citenamefont {Felvinti}, \citenamefont
  {Winward}, \citenamefont {Dixon}, \citenamefont {Minder}, \citenamefont
  {Kumar} \emph {et~al.}}]{amies2023quantum}%
  \BibitemOpen
  \bibfield  {author} {\bibinfo {author} {\bibfnamefont {B.}~\bibnamefont
  {Amies-King}}, \bibinfo {author} {\bibfnamefont {K.~P.}\ \bibnamefont
  {Schatz}}, \bibinfo {author} {\bibfnamefont {H.}~\bibnamefont {Duan}},
  \bibinfo {author} {\bibfnamefont {A.}~\bibnamefont {Biswas}}, \bibinfo
  {author} {\bibfnamefont {J.}~\bibnamefont {Bailey}}, \bibinfo {author}
  {\bibfnamefont {A.}~\bibnamefont {Felvinti}}, \bibinfo {author}
  {\bibfnamefont {J.}~\bibnamefont {Winward}}, \bibinfo {author} {\bibfnamefont
  {M.}~\bibnamefont {Dixon}}, \bibinfo {author} {\bibfnamefont
  {M.}~\bibnamefont {Minder}}, \bibinfo {author} {\bibfnamefont
  {R.}~\bibnamefont {Kumar}}, \emph {et~al.},\ }\bibfield  {title} {\bibinfo
  {title} {Quantum communications feasibility tests over a uk-ireland 224 km
  undersea link},\ }\href@noop {} {\bibfield  {journal} {\bibinfo  {journal}
  {Entropy}\ }\textbf {\bibinfo {volume} {25}},\ \bibinfo {pages} {1572}
  (\bibinfo {year} {2023})}\BibitemShut {NoStop}%
\bibitem [{\citenamefont {Zanforlin}\ \emph {et~al.}(2022)\citenamefont
  {Zanforlin}, \citenamefont {Lupo}, \citenamefont {Connolly}, \citenamefont
  {Kok}, \citenamefont {Buller},\ and\ \citenamefont
  {Huang}}]{zanforlin2022optical}%
  \BibitemOpen
  \bibfield  {author} {\bibinfo {author} {\bibfnamefont {U.}~\bibnamefont
  {Zanforlin}}, \bibinfo {author} {\bibfnamefont {C.}~\bibnamefont {Lupo}},
  \bibinfo {author} {\bibfnamefont {P.~W.}\ \bibnamefont {Connolly}}, \bibinfo
  {author} {\bibfnamefont {P.}~\bibnamefont {Kok}}, \bibinfo {author}
  {\bibfnamefont {G.~S.}\ \bibnamefont {Buller}},\ and\ \bibinfo {author}
  {\bibfnamefont {Z.}~\bibnamefont {Huang}},\ }\bibfield  {title} {\bibinfo
  {title} {Optical quantum super-resolution imaging and hypothesis testing},\
  }\href@noop {} {\bibfield  {journal} {\bibinfo  {journal} {Nature
  Communications}\ }\textbf {\bibinfo {volume} {13}},\ \bibinfo {pages} {5373}
  (\bibinfo {year} {2022})}\BibitemShut {NoStop}%
\bibitem [{\citenamefont {Notaros}\ \emph {et~al.}(2017)\citenamefont
  {Notaros}, \citenamefont {Mower}, \citenamefont {Heuck}, \citenamefont
  {Lupo}, \citenamefont {Harris}, \citenamefont {Steinbrecher}, \citenamefont
  {Bunandar}, \citenamefont {Baehr-Jones}, \citenamefont {Hochberg},
  \citenamefont {Lloyd},\ and\ \citenamefont {Englund}}]{Notaros:17}%
  \BibitemOpen
  \bibfield  {author} {\bibinfo {author} {\bibfnamefont {J.}~\bibnamefont
  {Notaros}}, \bibinfo {author} {\bibfnamefont {J.}~\bibnamefont {Mower}},
  \bibinfo {author} {\bibfnamefont {M.}~\bibnamefont {Heuck}}, \bibinfo
  {author} {\bibfnamefont {C.}~\bibnamefont {Lupo}}, \bibinfo {author}
  {\bibfnamefont {N.~C.}\ \bibnamefont {Harris}}, \bibinfo {author}
  {\bibfnamefont {G.~R.}\ \bibnamefont {Steinbrecher}}, \bibinfo {author}
  {\bibfnamefont {D.}~\bibnamefont {Bunandar}}, \bibinfo {author}
  {\bibfnamefont {T.}~\bibnamefont {Baehr-Jones}}, \bibinfo {author}
  {\bibfnamefont {M.}~\bibnamefont {Hochberg}}, \bibinfo {author}
  {\bibfnamefont {S.}~\bibnamefont {Lloyd}},\ and\ \bibinfo {author}
  {\bibfnamefont {D.}~\bibnamefont {Englund}},\ }\bibfield  {title} {\bibinfo
  {title} {Programmable dispersion on a photonic integrated circuit for
  classical and quantum applications},\ }\href
  {https://doi.org/10.1364/OE.25.021275} {\bibfield  {journal} {\bibinfo
  {journal} {Opt. Express}\ }\textbf {\bibinfo {volume} {25}},\ \bibinfo
  {pages} {21275} (\bibinfo {year} {2017})}\BibitemShut {NoStop}%
\bibitem [{\citenamefont {Politi}\ \emph {et~al.}(2009)\citenamefont {Politi},
  \citenamefont {Matthews}, \citenamefont {Thompson},\ and\ \citenamefont
  {O'Brien}}]{politi2009integrated}%
  \BibitemOpen
  \bibfield  {author} {\bibinfo {author} {\bibfnamefont {A.}~\bibnamefont
  {Politi}}, \bibinfo {author} {\bibfnamefont {J.~C.}\ \bibnamefont
  {Matthews}}, \bibinfo {author} {\bibfnamefont {M.~G.}\ \bibnamefont
  {Thompson}},\ and\ \bibinfo {author} {\bibfnamefont {J.~L.}\ \bibnamefont
  {O'Brien}},\ }\bibfield  {title} {\bibinfo {title} {Integrated quantum
  photonics},\ }\href@noop {} {\bibfield  {journal} {\bibinfo  {journal} {IEEE
  Journal of Selected Topics in Quantum Electronics}\ }\textbf {\bibinfo
  {volume} {15}},\ \bibinfo {pages} {1673} (\bibinfo {year}
  {2009})}\BibitemShut {NoStop}%
\bibitem [{\citenamefont {Moody}\ \emph {et~al.}(2022)\citenamefont {Moody},
  \citenamefont {Sorger}, \citenamefont {Blumenthal}, \citenamefont
  {Juodawlkis}, \citenamefont {Loh}, \citenamefont {Sorace-Agaskar},
  \citenamefont {Jones}, \citenamefont {Balram}, \citenamefont {Matthews},
  \citenamefont {Laing} \emph {et~al.}}]{moody20222022}%
  \BibitemOpen
  \bibfield  {author} {\bibinfo {author} {\bibfnamefont {G.}~\bibnamefont
  {Moody}}, \bibinfo {author} {\bibfnamefont {V.~J.}\ \bibnamefont {Sorger}},
  \bibinfo {author} {\bibfnamefont {D.~J.}\ \bibnamefont {Blumenthal}},
  \bibinfo {author} {\bibfnamefont {P.~W.}\ \bibnamefont {Juodawlkis}},
  \bibinfo {author} {\bibfnamefont {W.}~\bibnamefont {Loh}}, \bibinfo {author}
  {\bibfnamefont {C.}~\bibnamefont {Sorace-Agaskar}}, \bibinfo {author}
  {\bibfnamefont {A.~E.}\ \bibnamefont {Jones}}, \bibinfo {author}
  {\bibfnamefont {K.~C.}\ \bibnamefont {Balram}}, \bibinfo {author}
  {\bibfnamefont {J.~C.}\ \bibnamefont {Matthews}}, \bibinfo {author}
  {\bibfnamefont {A.}~\bibnamefont {Laing}}, \emph {et~al.},\ }\bibfield
  {title} {\bibinfo {title} {2022 roadmap on integrated quantum photonics},\
  }\href@noop {} {\bibfield  {journal} {\bibinfo  {journal} {Journal of
  Physics: Photonics}\ }\textbf {\bibinfo {volume} {4}},\ \bibinfo {pages}
  {012501} (\bibinfo {year} {2022})}\BibitemShut {NoStop}%
\bibitem [{\citenamefont {Wang}\ \emph
  {et~al.}(2019{\natexlab{b}})\citenamefont {Wang}, \citenamefont {Qin},
  \citenamefont {Ding}, \citenamefont {Chen}, \citenamefont {Chen},
  \citenamefont {You}, \citenamefont {He}, \citenamefont {Jiang}, \citenamefont
  {You}, \citenamefont {Wang}, \citenamefont {Schneider}, \citenamefont
  {Renema}, \citenamefont {H\"ofling}, \citenamefont {Lu},\ and\ \citenamefont
  {Pan}}]{PhysRevLett.123.250503}%
  \BibitemOpen
  \bibfield  {author} {\bibinfo {author} {\bibfnamefont {H.}~\bibnamefont
  {Wang}}, \bibinfo {author} {\bibfnamefont {J.}~\bibnamefont {Qin}}, \bibinfo
  {author} {\bibfnamefont {X.}~\bibnamefont {Ding}}, \bibinfo {author}
  {\bibfnamefont {M.-C.}\ \bibnamefont {Chen}}, \bibinfo {author}
  {\bibfnamefont {S.}~\bibnamefont {Chen}}, \bibinfo {author} {\bibfnamefont
  {X.}~\bibnamefont {You}}, \bibinfo {author} {\bibfnamefont {Y.-M.}\
  \bibnamefont {He}}, \bibinfo {author} {\bibfnamefont {X.}~\bibnamefont
  {Jiang}}, \bibinfo {author} {\bibfnamefont {L.}~\bibnamefont {You}}, \bibinfo
  {author} {\bibfnamefont {Z.}~\bibnamefont {Wang}}, \bibinfo {author}
  {\bibfnamefont {C.}~\bibnamefont {Schneider}}, \bibinfo {author}
  {\bibfnamefont {J.~J.}\ \bibnamefont {Renema}}, \bibinfo {author}
  {\bibfnamefont {S.}~\bibnamefont {H\"ofling}}, \bibinfo {author}
  {\bibfnamefont {C.-Y.}\ \bibnamefont {Lu}},\ and\ \bibinfo {author}
  {\bibfnamefont {J.-W.}\ \bibnamefont {Pan}},\ }\bibfield  {title} {\bibinfo
  {title} {Boson sampling with 20 input photons and a 60-mode interferometer in
  a $1{0}^{14}$-dimensional hilbert space},\ }\href
  {https://doi.org/10.1103/PhysRevLett.123.250503} {\bibfield  {journal}
  {\bibinfo  {journal} {Phys. Rev. Lett.}\ }\textbf {\bibinfo {volume} {123}},\
  \bibinfo {pages} {250503} (\bibinfo {year} {2019}{\natexlab{b}})}\BibitemShut
  {NoStop}%
\end{thebibliography}
\end{document}